\documentclass[preprint,aps,nofootinbib,preprintnumbers]{revtex4-1}
\pdfoutput=1

\usepackage{amsmath,amssymb,amsfonts,dcolumn,color,graphicx,graphics,latexsym,placeins,epsfig}
\usepackage{epsfig}
\usepackage{bm}
\usepackage{slashed}
\usepackage{latexsym}
\usepackage{natbib}
\usepackage{url}
\usepackage{dcolumn}
\usepackage{color}
\usepackage{amsfonts,amssymb,amsmath}
\usepackage{graphicx,epsfig}
\usepackage{psfrag}
\usepackage{tabularx}
\usepackage{hyperref}
\hypersetup{colorlinks=true}
\usepackage{comment}
\usepackage{graphicx}

\usepackage{ragged2e}
\usepackage{subcaption}
\DeclareCaptionJustification{justified}{\justifying}

\begin{document}
  
\preprint{YITP-25-168, RESCEU-22/25, IPMU25-0050}

\title{Exorcising ghosts with gravitational waves: cases of ghostful and ghost-free fourth-order gravity}

\author{Gaetano Lambiase $^{1,2}$\footnote{lambiase@sa.infn.it}}
\author{Shinji Mukohyama$^{3,4,5}$\footnote{shinji.mukohyama@yukawa.kyoto-u.ac.jp}}
\author{Tanmay Kumar Poddar$^{2}$\footnote{poddar@sa.infn.it}}
\author{Anna Chiara Rescigno$^{1}$\footnote{annachiara.r@gmail.com}}

\affiliation{$^{1}$Dipartimento di Fisica ``E.R. Caianiello'' Universit\`a di Salerno, I-84084 Fisciano (Sa), Italy,}
\affiliation{$^{2}$ INFN, Gruppo collegato di Salerno, Via Giovanni Paolo II 132, Fisciano I-84084, Italy}
\affiliation{$^{3}$ Center for Gravitational Physics and Quantum Information, Yukawa Institute for Theoretical Physics,
Kyoto University, 606-8502, Kyoto, Japan}
\affiliation{$^{4}$ Research Center for the Early Universe (RESCEU), Graduate School of Science, The University of Tokyo, Hongo 7-3-1, Bunkyo-ku, Tokyo 113-0033, Japan}
\affiliation{$^{5}$ Kavli Institute for the Physics and Mathematics of the Universe (WPI), The University of Tokyo Institutes for Advanced Study (UTIAS), The University of Tokyo, Kashiwa, Chiba 277-8583, Japan}

\begin{abstract}
General Relativity (GR) is an effective field theory valid in the infrared regime. Quadratic curvature extensions intended to probe ultraviolet physics generically propagate a massive spin-$2$ ghost and are therefore non-unitary. One route to remove ghost is by enlarging the geometric sector (torsion, non-metricity). We investigate the infrared phenomenology of both the standard (ghostful) and ghost-free fourth-order gravity theories by computing Gravitational Wave (GW) emission and confronting the results with observations such as the orbital-period decay of quasi-stable binaries such as PSR B1913+16 and PSR J1738+0333 and the chirp-mass evolution of GW170817. In the ghostful theory, besides the theoretical inconsistency due to non-unitarity, there are also phenomenological problems: the massless spin-$2$ GW flux cancels the combined GW fluxes of the massive spin-$2$ ghost and massive spin-$0$ scalar in the vanishing-mass limit, so the GR quadrupole formula is not recovered at the leading order. 
As a result, we obtain the GW constraint on the ghostful theory as $m\gtrsim 10^{-11}~\mathrm{eV}$, where $m$ is the mass of the massive modes. By contrast, the ghost-free theory smoothly reproduces the Newtonian potential and GR quadrupole formulae when the two coupling constants $\alpha_1$ and $\alpha_2$ vanish, independently of the mass $m$. Therefore, GW observations put mass-dependent upper bounds on the size of the coupling constants. For example, if we assume $\alpha_1\simeq\alpha_2$ for simplicity, then we obtain $\alpha_{1,2}\lesssim 4.2\times 10^{83}$ for $m\sim 3\times 10^{-16}\,\mathrm{eV}$ and $\alpha_{1,2}\lesssim 1.3\times 10^{75}$ for $m\sim 10^{-11}\,\mathrm{eV}$. To our knowledge, these are the first astrophysical-scale bounds reported for ghostful and ghost-free fourth-order gravity.
\end{abstract}

\maketitle
\section{Introduction}\label{sec1}

Since its formulation, General Relativity (GR) has been subjected to numerous experimental tests, all of which have shown remarkable agreement with observations in the InfraRed (IR) regime, i.e., at large distances and late times \cite{Will:2014kxa}. The indirect evidence for Gravitational Waves (GWs) from pulsar timing in systems such as the Hulse-Taylor (HT) binary PSR B1913+16 \cite{Hulse:1974eb}, together with the direct detections of GW emission from compact binary mergers-most notably GW150914 (Black Hole-Black Hole (BH-BH)) \cite{LIGOScientific:2016aoc} and GW170817 (Neutron Star-Neutron Star (NS-NS)) \cite{LIGOScientific:2017zic} observed by LIGO-Virgo-further confirm the consistency of GR with experimental data.

Despite its remarkable success in the IR regime, GR faces open challenges in the UltraViolet (UV) domain. At the classical level, cosmological and BH singularities remain unresolved, while at the quantum level the theory suffers from non-renormalizability, posing a major obstacle to its consistency.

Another open issue concerns the behavior of gravity at short distances. The Newtonian $1/r$ potential has been experimentally tested only down to scales of about $10~\mu\mathrm{m}$ in torsion-balance experiments \cite{Kapner:2006si}, corresponding to energies of order $10^{-2}~\mathrm{eV}$. Beyond this scale, our understanding of gravity is purely theoretical, and extrapolating GR all the way to the Planck scale ($M_{pl} \sim 10^{19}~\mathrm{GeV}$) remains speculative.

The limitations of GR in the UV regime suggest that the GR should be regarded as a low-energy Effective Field Theory (EFT). Over the years, various approaches have been proposed to unify gravity with quantum theory, including superstring theory \cite{Schwarz:2000ew}, Ho\v{r}ava-Lifshitz gravity \cite{Sotiriou:2010wn}, and ghost-free nonlocal gravity \cite{Nojiri:2019dio}, among others. While the construction of a consistent quantum gravity theory is a fundamental objective, it is equally important to investigate its possible low-energy manifestations. This can be done either through quantum corrections to GR within the EFT framework or, more broadly, through extended theories of gravity.

A well-known example is the Starobinsky model \cite{Starobinsky:1980te}, which generalizes the Einstein-Hilbert action by adding an $R^2$ term, where $R$ denotes the Ricci scalar, leading to profound implications for inflationary cosmology \cite{Planck:2018jri, Hazumi:2019lys}. Another natural extension of Einstein's theory is to couple scalar fields to curvature invariants or to include higher-order curvature terms, such as $R^2$, $R_{\mu\nu}R^{\mu\nu}$, and $R_{\mu\nu\rho\sigma}R^{\mu\nu\rho\sigma}$, where $R_{\mu\nu}$ is the Ricci tensor, and $R_{\mu\nu\rho\sigma}$ is the Riemann curvature tensor.

The inclusion of higher-curvature terms in the action introduces higher derivatives of the graviton field, which in turn give rise to a massive spin-$2$ Ostrogradsky ghost. As first demonstrated in the seminal work of Stelle \cite{Stelle:1976gc}, a purely quadratic curvature action in four dimensions is power-counting renormalizable; however, it is non-unitary due to the propagation of a spin-$2$ ghost degree of freedom. This tension between renormalizability and unitarity poses a major obstacle, suggesting that a consistent perturbative theory of quantum gravity may be unattainable. One proposed resolution involves nonlocal extensions of gravity, in which nonlocality is introduced via form factors containing an infinite series of covariant derivatives \cite{Krasnikov:1987yj,Tomboulis:1997gg,Tseytlin:1995uq,Siegel:2003vt,Biswas:2005qr,Biswas:2011ar,Biswas:2016etb,Biswas:2016egy}. Another possibility is to maintain the Euclidean signature of the metric at the fundamental level and to invoke emergence of the Lorentz signature and invariance only at long distances \cite{Mukohyama:2013ew,Mukohyama:2013gra,Muneyuki:2013aba,Kehayias:2014uta,Feng:2023klt,Feng:2025xsi}. 

If the ghost contribution becomes dominant, the theory lies outside the domain of validity of the EFT, leading to a loss of theoretical control. Predictability can only be maintained if the ghost mass (when present) is much larger than the relevant energy scale. In the special case where the ghost mass associated with generic higher-curvature terms is pushed to infinity, the theory effectively reduces to an $f(R)$ model, which is ghost-free. Moreover, $f(R)$ theories can be reformulated equivalently as scalar-tensor theories~\cite{Fujii:2003pa}.

In \cite{Aoki:2019snr}, it was shown that derivative corrections to Einstein gravity can generate, in addition to the masless spin-$2$ mode, a massive spin-$2$ particle, a massive spin-$0$ particle, without introducing light ghosts. The key observation of \cite{Aoki:2019snr} is that, from a geometric standpoint, the metric and the connection may be treated as independent variables. Under specific conditions on the coupling constants, this framework admits new classes of theories in which the massless spin-$2$ graviton, a massive spin-$2$ mode, and a massive spin-$0$ mode -- similar to the spectrum in quadratic curvature gravity -- can coexist without ghost instabilities, at least around the Minkowski background.

It is worth noting that these new classes of higher-curvature theories are closely connected to recent advances in modified gravity, where several traditional no-go results for ghost-free constructions have been circumvented. Prominent examples include ghost-free massive gravity \cite{deRham:2010ik,deRham:2010kj} and ghost-free scalar–tensor frameworks with higher-derivative interactions, such as the Degenerate Higher-Order Scalar-Tensor (DHOST) theories \cite{Gleyzes:2014dya,Gleyzes:2014qga,Zumalacarregui:2013pma,Langlois:2015cwa,Langlois:2015skt,Crisostomi:2016czh,BenAchour:2016cay,BenAchour:2016fzp}. Comprehensive treatments of GR within the EFT framework and its phenomenological signatures have been studied in \cite{Donoghue:1994dn,Porto:2016pyg,Cardoso:2018ptl,Sennett:2019bpc}.

In this paper, we compute the rate of energy loss due to GW radiation in two frameworks, the standard or ghostful fourth-order gravity, which contains a massless spin-$2$ graviton, a massive spin-$2$ ghost, and a massive spin-$0$ scalar; and the ghost-free quadratic gravity, which features the same spectrum except that the massive spin-$2$ mode is non-ghostlike. The presence of these additional modes modifies the gravitational potential from its standard Newtonian form $(1/r)$; in particular, the potential in standard fourth-order gravity becomes non-singular at $r=0$.

The total GW energy loss receives contributions from both the modified force law and from the additional radiation channels associated with the massive modes. We study quasi-stable binary systems such as the HT binary (pulsar-NS) and PSR J1738+0333 (NS-White Dwarf (WD)) \cite{Freire:2012mg}, where the observed orbital period decay provides indirect confirmation of GW emission through pulsar timing. Note, the HT binary agrees with Einstein's GR prediction for GW radiation at the sub-percent level ($<0.1\%$) \cite{Hulse:1974eb,taylor1982new,Weisberg:1984zz,Weisberg:2016jye}.

The direct detection of GWs has been achieved from coalescing binaries, beginning with the GW150914 event (a binary BH merger) and later with the GW170817 event (a binary NS merger). Unlike quasi-stable binaries, coalescing systems have time-varying orbital frequencies and shrinking separations, yet the measured chirp mass, GW strain, and phase from these GW events also confirm GR.

Similar analyses have been carried out in the context of ultralight bosonic particles, which could be well-motivated Dark Matter (DM) candidates. Radiation of such particles can contribute to the energy loss in both quasi-stable and coalescing binaries, and observational data can thus be used to place constraints on their properties \cite{KumarPoddar:2019ceq,KumarPoddar:2019jxe,Dror:2019uea}. If the stars carry dark charges, an additional fifth force arises alongside gravity \cite{Kopp:2018jom,Poddar:2023pfj}. Moreover, several massive gravity scenarios-including the Fierz-Pauli theory \cite{Visser:1997hd,Gambuti:2020onb,Gambuti:2021meo}, the DGP model \cite{Dvali:2000hr,Dvali:2000rv,Dvali:2000xg}, and modified FP extensions \cite{Finn:2001qi} have also been tested against GW observations \cite{Poddar:2021yjd}.

The existence of massive modes in fourth-order gravity can alter both the orbital period loss in quasi-stable binaries and the frequency evolution in coalescing binaries. Using the observed orbital period decay of the HT and PSR J1738+0333 systems, we obtain constraints on the masses of the massive ghost and scalar modes in standard fourth-order gravity. In the ghost-free variant, we constrain instead the masses of the massive spin-$2$ and scalar modes along with their couplings. The bounds on the mass and coupling parameters in both theories are also obtained from the GW170817 binary NS merger. Specifically, the contribution from the modified force law becomes relevant for mode masses $m \lesssim 1/a \sim 10^{-16}~\mathrm{eV}$ in quasi-stable binaries, while the radiation channel contributes for $m \lesssim \Omega \sim 10^{-19}~\mathrm{eV}$, where $a$ and $\Omega$ characterize the separation between the two stars and the orbital frequency of the binary respectively. For coalescing binaries, the corresponding thresholds are $m \lesssim 1/a \sim 10^{-11}~\mathrm{eV}$ for the new force contribution and $m \lesssim \Omega \sim 10^{-14}~\mathrm{eV}$ for radiation, where $a$ corresponds to approximately the two times the radius of the NS and $\Omega$ is the orbital frequency of the binary when it just enters the LIGO frequency band. In the ghost-free quadratic gravity framework, the bounds on the couplings depend explicitly on these mass parameters.

The paper is structured as follows. In section \ref{sec2}, we provide a brief overview of ghost-free quadratic gravity. In section \ref{sec4}, we derive the gravitational potential in the ghost and non-ghost versions of fourth-order gravity. In section \ref{sec3}, we compute the rate of GW energy loss in both standard fourth-order gravity and ghost-free quadratic gravity, including contributions from radiation due to massive modes. Section \ref{sec5} is devoted to the orbital period loss in quasi-stable binaries, where both the modified force law and additional radiation channels contribute. In section  \ref{sec6}, we analyze the orbital frequency evolution in coalescing binaries within the same frameworks. Constraints on the masses of the modes and the coupling parameters are then extracted from the orbital period decay of the HT and PSR J1738+0333 systems in section \ref{sec7}, and from the GW170817 event in section \ref{sec8}. Finally, section \ref{sec9} summarizes our findings and presents our conclusions.

We use natural system of units $c$ (speed of light)$= \hbar$ (reduced Planck constant)$=1$ throughout the paper, with the reduced Planck mass defined as $M_{pl}^2 = 1/(8\pi G)$, unless stated otherwise.

\section{Overview on ghost-free quadratic gravity}\label{sec2}

In this section, we review the generalized theory of fourth-order gravity as proposed in \cite{Aoki:2019snr}, which explores gravity models incorporating derivative corrections to GR. The focus is on identifying new classes of ghost-free theories formulated around a Minkowski background. The framework assumes a Riemann-Cartan geometry, allowing for non-zero torsion and treating the metric and affine connection (with vanishing non-metricity) as independent variables. This generalization introduces additional massive modes associated with torsion, and leads to modifications in the Lagrangian that help eliminate ghost instabilities. The most general parity-preserving Lagrangian including all terms up to mass dimension four has been considered. It encompasses the Einstein-Hilbert (EH) term, quadratic curvature invariants, as well as contributions involving torsion and its derivatives as
\begin{equation}
\mathcal{L}=\mathcal{L}_{EH}+\mathcal{L}_{R^2}+\mathcal{L}_{T^2} + \mathcal{L}_{(\nabla T)^2}+\mathcal{L}_{R\nabla T} + \mathcal{L}_{RT^2}+\mathcal{L}_{(\nabla T)T^2}+\mathcal{L}_{T^4},
\label{s1}
\end{equation}
where $T$ represents the torsion tensor. The last three terms are of higher-order in the perturbative analysis around the Minkowski background, while the first two terms are given as
\begin{equation}
\mathcal{L}_{EH}=\frac{M_{pl}^2}{2} R, \hspace{0.5cm} \mathcal{L}_{R^2}=\alpha_1 R^2 +\alpha_2 R_{\mu\nu}R^{\mu\nu}+\alpha_3 R_{\mu\nu\rho\sigma}R^{\mu\nu\rho\sigma},
\label{s2}
\end{equation}
where the derivative correction to the Einstein-Hilbert Lagrangian, denoted by $\mathcal{L}_{EH}$, is captured by the term $\mathcal{L}_{R^2}$, which arises from quantum corrections. In four dimensions, one of the coefficients $\alpha_i$ can be eliminated using the Gauss-Bonnet theorem. Hereafter, we use this freedom to set $\alpha_3$ to zero. In extended gravity theories beyond fourth-order, additional higher-derivative terms may also appear in $\mathcal{L}_{R^2}$ (as well as in other terms).

The mass term of the torsion field is given as
\begin{equation}
\mathcal{L}_{T^2}=\frac{M_T^2}{2}(a_1 \overset{\scriptscriptstyle  (1)}{T } {}^{\mu\nu\rho}\overset{\scriptscriptstyle  (1)}{T } {}_{\mu\nu\rho}+a_2  T^{\mu}T_{\mu}+a_3\mathcal{T}^{\mu} \mathcal{T}_{\nu}),
\label{s3}
\end{equation}
where the coefficients $a_i$ must be non-zero to ensure that standard GR is recovered in the low-energy limit and the three irreducible torsion tensor components are
\begin{equation}
\overset{\scriptscriptstyle  (1)}{T } {}_{\mu\nu\rho}=T_{\mu\nu\rho}-\overset{\scriptscriptstyle  (2)}{T }{}_{\mu\nu\rho}-\overset{\scriptscriptstyle  (3)}{T }{}_{\mu\nu\rho},~~~\overset{\scriptscriptstyle  (2)}{T } {} _{\mu\nu\rho} =\frac{2}{3}g_{\mu[\nu}T_{\rho]},~~~\overset{\scriptscriptstyle  (3)}{T } {}_{\mu\nu\rho}=\epsilon_{\mu\nu\rho\sigma}\mathcal{T}^{\sigma}, 
\label{s6}
\end{equation}
where
\begin{equation}
T_{\mu}:=T^{\nu}{}_{\nu\mu}, ~~~\mathcal{T}_{\mu}:=\frac{1}{6}\epsilon_{\mu\nu\rho\sigma}T^{\nu\rho\sigma}, 
\label{s7}
\end{equation}
with the irreducible piece $\overset{\scriptscriptstyle  (1)}{T } {}_{\mu\nu\rho}$ satisfying
\begin{equation}
\overset{\scriptscriptstyle  (1)}{T } {}_{\mu(\nu\rho)}=0,\quad \overset{\scriptscriptstyle  (1)}{T } {}_{[\mu\nu\rho]}=0, \quad \overset{\scriptscriptstyle  (1)}{T } {}^{\mu}{}_{\mu\nu}=0.  
\label{s8}
\end{equation}

The source term for the torsion equation of motion is governed by the Lagrangian
\begin{equation}
\mathcal{L}_{R\nabla T}=\beta_1 R_{\mu\nu} \nabla_{\rho}\overset{\scriptscriptstyle  (1)}{T } {}^{\mu\nu\rho}+\beta_2 R \nabla_{\mu} T^{\mu},
\label{s4}
\end{equation}
where in the limit $\beta_1 = \beta_2 = 0$, the metric and torsion fields decouple at linear order around Minkowski spacetime. However, in this limit, the massive spin-2 ghost persists as long as $\alpha_2 \ne 0$.

The remaining relevant term in Eq. \ref{s1} is
\begin{equation}
\begin{split}
\mathcal{L}_{(\nabla T)^2}=b_1 \nabla_{\mu} \overset{\scriptscriptstyle  (1)}{T } {}_{\nu\rho\sigma}\nabla^{\mu} \overset{\scriptscriptstyle  (1)}{T } {}^{\nu\rho\sigma}+b_2 \nabla_{\mu} \overset{\scriptscriptstyle  (1)}{T } {}^{\mu\rho\sigma} \nabla_{\nu} \overset{\scriptscriptstyle  (1)}{T } {}^{\nu}{}_{\rho\sigma}+ b_3 \nabla_{\mu} \overset{\scriptscriptstyle  (1)}{T } {}^{\rho\sigma\mu} \nabla_{\nu} \overset{\scriptscriptstyle  (1)}{T } {}_{\rho\sigma}{}^{\nu}
+b_4 \nabla_{\mu}T_{\nu} \nabla^{\mu} T^{\nu}+\\
b_5 \nabla_{\mu} T^{\mu} \nabla_{\nu} T^{\nu}
+b_6 \nabla_{\mu}\mathcal{T}_{\nu} \nabla^{\mu} \mathcal{T}^{\nu}  +b_7 \nabla_{\mu} \mathcal{T}^{\mu} \nabla_{\nu} \mathcal{T}^{\nu}+b_8\nabla_{\mu} \overset{\scriptscriptstyle  (1)}{T } {}^{\mu\nu\rho}\nabla_{\nu}T_{\rho} 
+b_9 \epsilon^{\mu\nu\rho\sigma} \nabla_{\alpha} \overset{\scriptscriptstyle  (1)}{T ^{\alpha}}{}_{\mu\nu} \nabla_{\rho}\mathcal{T}_{\sigma},   
\end{split}
\label{s5}
\end{equation}
where $b_i$ along with $\alpha_i,~\beta_i, ~a_i$ are all dimensionless parameters.

Thus, the linear perturbations of the theory around the Minkowski background are characterized by a total of 16 dimensionless parameters $(\alpha_i, \beta_i, a_i, b_i)$, along with two mass scales: the torsion mass $M_T$ and the Planck mass $M_{pl}$. In the limit where the torsion mass $M_T$ becomes infinitely large, while keeping $b_i$ and $\beta_i$ finite, all torsional degrees of freedom decouple and can be integrated out, reducing the theory effectively to the purely metric formulation.

In the limit where the canonical spin tensor vanishes, the interaction Lagrangian between matter and gravity reduces to $(1/2)\, \delta g_{\mu\nu} T^{\mu\nu}$. In this framework, the propagating gravitational modes consist of a massless spin-2 graviton, a massive spin-2 mode, and a massive spin-0 scalar.

The matrix form for the kinetic matrices $K_{2^+}$ and $K_{0^+}$ in the momentum space for the spin-$2$ and spin-$0$ degrees of freedom in the ghost-free fourth-order gravity, which arise from the perturbative expansion of the Lagrangian in Eq. \ref{s1}, is given as \cite{Aoki:2019snr}
\begin{align}
K_{2^+}&:=
-k^2 
\begin{pmatrix}
\frac{1}{4}M_{pl}^2-\frac{1}{2}\alpha_2 k^2 & -\frac{1}{2}\beta_1 k^2 \\
* & -2a_1 M_T^2-2(2b_1+b_3)k^2
\end{pmatrix}
, \\
K_{0^+}& :=
-k^2
\begin{pmatrix}
-\frac{1}{2}M_{pl}^2 -2(3\alpha_1+\alpha_2)k^2 & \sqrt{3}\beta_2 k^2 \\
* &  -a_2 M_T^2-2(b_4+b_5)k^2
\end{pmatrix},
\end{align}
where the $*$ symbol indicates the symmetric parts of the matrices which we have omitted. The corresponding spin-2 and spin-0 projection operators, $P^{(2)}_{\mu\nu,\rho\sigma}$ and $P^{(0)}_{\mu\nu,\rho\sigma}$, are given by
\begin{equation}
P^{(2)}_{\mu\nu,\rho\sigma}=\theta_{\mu(\rho}\theta_{\sigma)\nu}-\frac{1}{3}\theta_{\mu\nu}\theta_{\rho\sigma},~~~P^{(0)}_{\mu\nu,\rho\sigma}=\frac{1}{3}\theta_{\mu\nu}\theta_{\rho\sigma},  
\label{s10}
\end{equation}
where 
\[
\theta_{\mu\nu}:=\eta_{\mu\nu}-(\partial_\mu \partial_\nu/\Box)\,. 
\]
The gauge-independent graviton propagator that enters the tree-level scattering amplitude is given by
\begin{equation}
-i\Delta_{\mu\nu,\rho\sigma} =i( K_{2^+}^{-1})^{11} P^{(2)}_{\mu\nu,\rho\sigma}+i(K_{0^+}^{-1})^{11}P^{(0)}_{\mu\nu,\rho\sigma},
\label{s9}
\end{equation}
where $(K_{2^+}^{-1})^{11}$ and $(K_{0^+}^{-1})^{11}$ denote the upper-left components of the inverse kinetic matrices $K_{2^+}^{-1}$ and $K_{0^+}^{-1}$, respectively.

In general the propagator has a $k^{-4}$ behavior in the high energy limit where $k$ is the internal momentum and 
\begin{eqnarray}
(K_{2^+}^{-1})^{11} &\rightarrow &  8\left(4\alpha_2-\frac{\beta_1^2}{2b_1+b_3}\right)^{-1} \frac{1}{k^4}+ \cdots \,,  \label{s11G} \\
( K_{0^+}^{-1})^{11} &\rightarrow & 2\left(4(3\alpha_1+\alpha_2)-\frac{3\beta_2^2}{b_4+b_5} \right)^{-1} \frac{1}{k^4}+\cdots, 
\label{s11}
\end{eqnarray}
where $\cdots$ correspond to $\mathcal{O}(k^{-6})$ and higher order terms. Note, Eqs. (\ref{s11G}, \ref{s11}) imply there are Ostrogradsky ghosts in the theory, and to eliminate the ghosts, the following critical conditions need to be satisfied as 
\begin{equation}
4\alpha_2(2b_1+b_3)-\beta_1^2=0, \qquad  4(3\alpha_1+\alpha_2)(b_4+b_5)-3\beta_2^2=0,   
\label{s12}
\end{equation}
with the propagator recovers the usual $k^{-2}$ behavior. Therefore, the complete propagator including all the propagating modes is obtained as
\begin{equation}
-i\Delta_{\mu\nu,\rho\sigma} = \frac{4}{M_{pl}^2} \mathcal{D}_{\mu\nu,\rho\sigma}
+\frac{8m_{2^+}^2\alpha_2 }{M_{pl}^4 } \frac{-i}{k^2+m_{2^+}^2} P^{(2)}_{\mu\nu,\rho\sigma}
+\frac{8m_{0^-}^2(3\alpha_1+\alpha_2)}{M_{pl}^4} \frac{-i}{k^2+m_{0^+}^2} P^{(0)}_{\mu\nu,\rho\sigma},    
\label{s13}
\end{equation}
where $\mathcal{D}_{\mu\nu,\rho\sigma}$ is the usual massless graviton propagator, given as
\begin{equation}
\mathcal{D}_{\mu\nu,\rho\sigma}=\frac{-i}{k^2}\left( \theta_{\mu(\rho}\theta_{\sigma)\nu}-\frac{1}{2}\theta_{\mu\nu}\theta_{\rho\sigma} \right),  
\label{s14}  
\end{equation}
and the masses of the massive spin-$2$ graviton mode $(m_{2+})$ and massive scalar spin-$0$ mode $(m_{0+})$ are
\begin{equation}
m_{2^+}^2 = \frac{4a_1 \alpha_2 M_{pl}^2 M_T^2}{M_{pl}^2 \beta_1^2-8a_1 \alpha_2^2 M_T^2},~~~ m_{0^+}^2 = \frac{2a_2(3\alpha_1+\alpha_2)M_{pl}^2 M_T^2}{8a_2 (3\alpha_1+\alpha_2)^2 M_T^2+3 \beta_2^2 M_{pl}^2 }.
\label{s15}  
\end{equation}

\vspace{1.0cm}\noindent {\bf Limit for standard fourth-order gravity:} In the limit of infinitely large torsion mass parameter $M_T$, the masses of the spin-$2$ and spin-$0$ modes, given by Eq. \ref{s15}, become 
\begin{equation}
m_{2^+}^2 \rightarrow -\frac{M_{pl}^2}{2\alpha_2} \quad \text{and} \quad 
m_{0^+}^2 \rightarrow \frac{M_{pl}^2}{4(3\alpha_1+\alpha_2)}\,,
\end{equation}
and the propagator in Eq. \ref{s13} is modified as
\begin{equation}
-i \Delta_{\mu\nu,\rho\sigma} \|_{M_T\rightarrow \infty} 
= \frac{4}{M_{pl}^2} \mathcal{D}_{\mu\nu,\rho\sigma}
-\frac{4}{M_{pl}^2 } \frac{-i}{q^2+m_{2^+}^2} P^{(2)}_{\mu\nu,\rho\sigma}+\frac{2}{M_{pl}^2} \frac{-i}{q^2+m_{0^+}^2} P^{(0)}_{\mu\nu,\rho\sigma},    
\label{s16}  
\end{equation}
which is the usual form of propagator in fourth-order gravity theories with massive spin-$2$ ghost.

\section{Gravitational potential}\label{sec4}

In the following, we calculate the gravitational potential for a non-relativistic source in ghost-free fourth-order gravity theory. (For the corresponding quantity in standard, i.e. ghostly, fourth-order gravity theory, see the last paragraph of this section.) The general tree-level scattering amplitude between two conserved source currents is given as
\begin{equation}
 \mathcal{M}=\frac{1}{4}T^{\mu\nu}(p_1,p_2) \Delta_{\mu\nu,\alpha\beta}(k) T^{\alpha\beta}(p_3,p_4),   
 \label{eq:1}
\end{equation}
where, $T^{\mu\nu}$ denotes the Fourier transform of the source current, with $p_i~(i=1,2,3,4)$ representing the external momenta, $k$ is the internal momentum and $\Delta_{\mu\nu,\alpha\beta}(k)$ denotes the gauge independent part of the graviton propagator.

For static sources, the saturated propagator in ghost-free quadratic gravity theory is constructed by contracting the propagator with the conserved external currents as
\begin{equation}
\mathcal{SP}(k)=T^{\mu\nu}(k)\Delta_{\mu\nu,\alpha\beta}(k)T^{\alpha\beta}(k)=\frac{A}{k^2}+\frac{B}{k^2+m_{2^+}^2}+\frac{C}{k^2+m_{0^+}^2},
\label{eq:2}
\end{equation}
where
\begin{equation}
A=\frac{4}{M_{pl}^2}\Big(|T_{\mu\nu}|^2-\frac{1}{2}T^2\Big),\quad
B= \frac{8m_{2^+}^2 \alpha_2}{M_{pl}^4}\Big(|T_{\mu\nu}|^2-\frac{1}{3}T^2\Big),\quad
C=\frac{8 m_{0^+}^2(3\alpha_1+\alpha_2)}{M_{pl}^4}\Big(\frac{1}{3}T^2\Big).
\label{eq:3}
\end{equation}
Since $A$, $B$, and $C$ are positive at the poles, the corresponding residues of the saturated propagator are positive as well, demonstrating the unitarity of the ghost-free fourth-order gravity theory in four dimensions within this framework. Thus, the propagating modes in these theories consist of a massless spin-2 tensor mode, a massive spin-2 tensor mode with mass $m^+_2$, and a massive spin-0 scalar mode with mass $m_0^+$ and all these modes are ghost-free. In deriving Eq.~\ref{eq:2}, we impose the conservation of the source current, $k_\mu T^{\mu\nu}=0$, which ensures that all momentum-dependent terms in the polarization sums for both massless and massive modes vanish.

For massive bodies at rest, the corresponding stress-energy tensors are $T^{\mu\nu} = (m_1, 0, 0, 0)$ and ${T'}^{\alpha\beta} = (m_2, 0, 0, 0)$. The massless and massive modes act as mediators in deriving the gravitational potential and hence, modifying the gravitational force. Under these conditions, the potential arising from the exchange of a massless graviton is given by
\begin{eqnarray}
V_A(r)&=&\frac{1}{4}\frac{4}{M^2_{pl}}\int \frac{d^3k}{(2\pi)^3}e^{i k\cdot r}\frac{1}{k^2}\Big(T_{\mu\nu}-\frac{1}{2}\eta_{\mu\nu}T^\alpha_\alpha\Big){T^\prime}^{\mu\nu}\nonumber\\
&=& \frac{m_1 m_2}{2M^2_{pl}}\int\frac{d^3k}{(2\pi)^3}\frac{e^{i\mathbf{k}\cdot \mathbf{r}}}{|\mathbf{k}|^2}=\frac{m_1 m_2}{2M^2_{pl}}\Big(\frac{1}{4\pi r}\Big)=\frac{Gm_1m_2}{r},
\label{pot1}
\end{eqnarray}
where in the second step, we apply the non-relativistic limit by assuming $k^0 \ll |\mathbf{k}|$. This yields the standard Newtonian potential mediated by a massless graviton between two massive objects. The potential arising from the exchange of a massive graviton between two conserved currents is given as
\begin{eqnarray}
V_B(r)&=&\frac{1}{4}\frac{8m^2_{2+}\alpha_2}{M^4_{pl}}\int \frac{d^3k}{(2\pi)^3}e^{ik\cdot r}\frac{1}{k^2+m^2_{2+}}\Big(T_{\mu\nu}-\frac{1}{3}\eta_{\mu\nu}T^\alpha_\alpha\Big){T^\prime}^{\mu\nu} \nonumber\\
&=& \frac{8\pi G}{4}\Big(\frac{8m^2_{2+}\alpha_2}{M^2_{pl}}\Big)\frac{2m_1m_2}{3}\frac{1}{4\pi r}e^{-m_{2+}r}=\frac{4}{3}\Big(\frac{2m^2_{2+}\alpha_2}{M^2_{pl}}\Big)\frac{Gm_1m_2}{r}e^{-m_{2+}r}.
\label{pot2}
\end{eqnarray}
Lastly, the potential arises from the exchange of a massive scalar between two conserved currents is obtained as
\begin{eqnarray}
V_C(r)&=&\frac{1}{4}\Big(\frac{8m^2_{0+}(3\alpha_1+\alpha_2)}{M^4_{pl}}\Big)\int \frac{d^3k}{(2\pi)^3}\frac{1}{k^2+m^2_{0+}}\Big(\frac{1}{3}\eta_{\mu\nu}T^\alpha_\alpha\Big){T^\prime}^{\mu\nu}\nonumber\\
&=& \frac{4}{3}\Big(\frac{(3\alpha_1+\alpha_2)m^2_{0+}}{M^2_{pl}}\Big)\frac{Gm_1m_2}{r}e^{-m_{0+}r},
\label{pot4}
\end{eqnarray}

In summary, the total potential for the fourth-order ghost-free gravity theory is 
\begin{equation}
V_{A}(r)+V_B(r)+V_C(r)=\frac{Gm_1m_2}{r}\Big[1+\frac{8}{3}\Big(\frac{m^2_{2+}\alpha_2}{M^2_{pl}}\Big)e^{-m_{2+}r}+\frac{4}{3}\Big(\frac{(3\alpha_1+\alpha_2)m^2_{0+}}{M^2_{pl}}\Big)e^{-m_{0+}r}\Big]. 
\label{pot6}
\end{equation}
Therefore, in the limit $m_{2+}, m_{0+}\rightarrow 0$, Eq. \ref{pot6} reduces to the standard Newtonian potential in the ghost-free quadratic gravity theory. However, the singularity at $r\rightarrow 0$ persists.

\vspace{1.0cm}\noindent {\bf Limit for standard fourth-order gravity:} In the limit $m^2_{2+}\rightarrow -M^2_{pl}/2\alpha_2$, Eq. \ref{pot2} reduces to
\begin{equation}
V_{B^\prime}(r)=-\frac{4}{3}\frac{Gm_1m_2}{r}e^{-m_{2+}r},    
\label{pot3}
\end{equation}
which corresponds to the potential mediated by the massive spin-2 ghost mode between two massive bodies. In the limit $m^2_{0+}\rightarrow M^2_{pl}/4(3\alpha_1+\alpha_2)$, Eq. \ref{pot4} becomes
\begin{equation}
V_{C^\prime}(r)=\frac{1}{3}\frac{Gm_1m_2}{r}e^{-m_{0+}r}. 
\label{pot5}
\end{equation}
Hence, the total potential for the standard fourth-order gravity is 
\begin{equation}
V_{A}(r)+V_{B^\prime}(r)+V_{C^\prime}(r)=\frac{Gm_1m_2}{r}\Big[1-\frac{4}{3}e^{-m_{2+}r}+\frac{1}{3}e^{-m_{0+}r}\Big].
\label{pot7}
\end{equation}
In the limit $m_{2+}, m_{0+}\rightarrow 0$, Eq. \ref{pot7} becomes finite at $r\rightarrow 0$ for the standard fourth-order gravity. Therefore, to obtain a non-zero potential in standard fourth-order gravity, we set $\alpha_2 =-M^2_{pl}/2m^2_{2+}\to 0$, effectively making the massive spin-2 ghost infinitely heavy. This leaves only the massless spin-2 graviton and a massive spin-0 scalar as the relevant propagating modes and the corresponding potential becomes
\begin{equation}
V_{A}(r)+V_{C^\prime}(r)=\frac{Gm_1m_2}{r}\Big[1+\frac{1}{3}e^{-m_{0+}r}\Big].
\end{equation}
This expression is valid for $m_{2+}r\gg 1$ ($\alpha_1\to 0$) and thus cannot be applied at $r=0$.

\section{Radiation of massless and massive modes}\label{sec3}

In this section, we evaluate the GW energy loss from a binary system due to the emission of both massless and massive modes in ghost-free fourth-order gravity theories, employing the Feynman diagram approach. (For the corresponding quantity in the standard, i.e. ghostly, fourth-order gravity theory, see subsection~\ref{subsec:radiation_standard4thordergravity}.) Here, the massive and massless modes act as on-shell bosonic particles. The binary system is treated as an effective one-body classical source in a center-of-mass frame, while the modes are described as quantum fields. Thus, we adopt a Quantum Field Theory (QFT) framework in the weak-field limit. In this picture, the computation of emission of the modes reduces to an effective single-vertex Feynman process. For standard GR, this method has been applied in \cite{Mohanty:1994yi,Poddar:2021yjd}, where it was shown that the QFT treatment in the weak-field approximation reproduces Einstein’s classical quadrupole radiation formula originally derived in \cite{Peters:1963ux} through multipole expansion method. Here, we extend the same formalism to both standard and ghost-free fourth-order gravity theories. Once the action is linearized and the propagator is determined, perturbative calculations are performed using tree-level Feynman diagrams. The emission rate of the massless and massive modes from the classical source follows from the interaction term in the Lagrangian, and involves the squared interaction amplitude, which depends on the source energy-momentum tensor ($T_{\mu\nu}$) and the polarization sum of the massless and massive modes which depend on the propagator structure of the theory.

The first $(A)$ term in Eq. \ref{eq:2} corresponds to the saturated propagator for the classical massless graviton emission in standard GR whereas the second $(B)$ and third $(C)$ terms correspond to saturated propagators for the emission of massive spin-2 tensor and massive spin-0 scalar modes. 

The emission rate for $i$'th $(i=A, B, C)$ mode is obtained as
\begin{equation}
d\Gamma_i= \frac{1}{4} |\mathcal{A}_i(k)|^2 2\pi \delta(\omega-\omega^\prime)\frac{d^3k}{(2\pi)^3}\frac{1}{2\omega},
\label{eq:4}
\end{equation}
where $\mathcal{A}_i(k)$ corresponds to the amplitude of the emission process for the $i$'th mode. Each propagating mode obeys a distinct dispersion relation. For the massless spin-$2$ graviton, the amplitude takes the form $\mathcal{A}_A \sim T_{\mu\nu}\epsilon^{\mu\nu}$, with $\epsilon_{\mu\nu}$ the polarization tensor; the corresponding squared amplitude is denoted by `$A$' in Eq. \ref{eq:3}. For the massive spin-$2$ mode, the structure is analogous, but the polarization sum and dispersion relation differ from the massless case, yielding the term `$B$' in Eq. \ref{eq:3}. The scalar mode instead couples to the trace of the energy-momentum tensor, and its squared amplitude is given by `$C$' in Eq. \ref{eq:3}, again with a modified dispersion relation entering the energy-loss integral.

Therefore, the rate of energy loss due to the radiation of $i$'th mode is
\begin{equation}
\Big(\frac{dE}{dt}\Big)_i=\frac{1}{8(2\pi)^2}\int |\mathcal{A}_i(k)|^2 \delta(\omega-\omega^\prime)k^2 dk d\Omega_k,
\label{eq:5}
\end{equation}
 where $d^3k=k^2dk d\Omega$. Hence, the total rate of energy loss due to the radiation of all modes are
\begin{equation}
\sum_i\Big(\frac{dE}{dt}\Big)_i=\Big(\frac{dE}{dt}\Big)_{A}+\Big(\frac{dE}{dt}\Big)_{B}+\Big(\frac{dE}{dt}\Big)_{C}, 
\label{eq:6}
\end{equation}
where 
\begin{eqnarray}
    \Big(\frac{dE}{dt}\Big)_{A} &=& \frac{1}{8(2\pi)^2}\int \frac{4}{M_{pl}^2}\Big[|T_{\mu\nu}(k')|^2-\frac{1}{2}|T^{\mu}{}_{\mu}(k^\prime)|^2\Big]\delta(\omega-\omega^\prime)\omega^2 d\omega d\Omega_k , \label{eq:7}\\
    \Big(\frac{dE}{dt}\Big)_{B}&=&\frac{1}{8(2\pi)^2}\int\frac{8m_{2^+}^2\alpha_2}{M_{pl}^4} \Big[|T_{\mu\nu}(k')|^2-\frac{1}{3}|T^{\mu}{}_{\mu}(k^\prime)|^2\Big]\delta(\omega-\omega^\prime)\omega^2 \Big(1-\frac{m^2_{2^+}}{\omega^2}\Big)^\frac{1}{2}d\omega d\Omega_k, \quad \label{eq:8}\\
    \Big(\frac{dE}{dt}\Big)_{C} &=&\frac{1}{8(2\pi)^2}\int \frac{8 m_{0^+}^2 (3\alpha_1+\alpha_2)}{M_{pl}^4}  \Big[\frac{1}{3}|T^{\mu}{}_{\mu}(k^\prime)|^2\Big]\delta(\omega-\omega^\prime)\omega^2 \Big(1-\frac{m^2_{0^+}}{\omega^2}\Big)^\frac{1}{2}d\omega d\Omega_k.
    \label{eq:9}
\end{eqnarray}
In the following, we compute the energy loss rate resulting from the emission of massless and massive gravitons, as well as massive scalar radiation.

\subsection{Rate of energy loss $(dE/dt)_A$ due to massless graviton radiation}\label{subseca}

We use the conserved current relation $k_\mu T^{\mu\nu}=0$, to obtain the stress-energy tensor components $T_{00}$ and $T_{i0}$ in terms of $T_{ij}$ as
\begin{equation}
 T_{00}=\hat{k^i}\hat{k^j}T_{ij}, \hspace{0.5cm}T_{0j}=-\hat{k^i}T_{ij}.
\label{eq:10}
\end{equation}
Therefore, we can write
 \begin{equation}
 \Big[|T_{\mu\nu}(k^\prime)|^2-\frac{1}{2}|T^{\mu}{}_{\mu}(k^\prime)|^2\Big]={\Lambda^A_{ij,lm}}T^{ij*}T^{lm},
\label{eq:11}
 \end{equation}
where the projection operator is defined as
\begin{equation}
{\Lambda^A_{ij,lm}}=\Big[\delta_{il}\delta_{jm}-2\hat{k_j}\hat{k_m}\delta_{il}+\frac{1}{2}\hat{k_i}\hat{k_j}\hat{k_l}\hat{k_m}-\frac{1}{2}\delta_{ij}\delta_{lm}+\frac{1}{2}\Big(\delta_{ij}\hat{k_l}\hat{k_m}+\delta_{lm}\hat{k_i}\hat{k_j}\Big)\Big].
\label{eq:12}
 \end{equation}
Now, using the relations 
 \begin{equation}
     \int d\Omega \hat{k^i}\hat{k^j}=\frac{4\pi}{3}\delta_{ij}, \hspace{0.5cm} \int d\Omega \hat{k^i}\hat{k^j}\hat{k^l}\hat{k^m}=\frac{4\pi}{15}(\delta_{ij}\delta_{lm}+\delta_{il}\delta_{jm}+\delta_{im}\delta_{jl}),
     \label{eq:13}
 \end{equation}
we perform the angular integral in Eq. \ref{eq:7} as
\begin{equation}
\int d\Omega_k \Lambda^A_{ij,lm}T^{ij*}(\omega^\prime)T^{lm}({\omega^\prime})=\frac{8\pi}{5}\Big(T_{ij}(\omega^\prime)T^*_{ji}(\omega^\prime)-\frac{1}{3}|T^{i}{}_{i}(\omega^\prime)|^2\Big),
\label{eq:14}
\end{equation}
where we use Eqs. \ref{eq:11} and \ref{eq:12}. 

The stress-energy tensor for a classical source corresponding to a compact binary system is given as
\begin{equation}
T_{\mu\nu}(x^\prime)=\mu \delta^3(\textbf{x}^\prime-\textbf{x}(t))U_\mu U_\nu,
\label{eq:15}
\end{equation}
where $\mu = \frac{m_1 m_2}{m_1 + m_2}$ denotes the reduced mass of the binary system, with $m_1$ and $m_2$ being the masses of the individual components. The four-velocity of the reduced mass in the non-relativistic limit is given by $U_\mu = (1, \dot{x}, \dot{y}, 0)$, corresponding to motion in the $x$-$y$ plane of the Keplerian orbit. The parametric form of the elliptic Keplerian orbit for a binary system is given as
\begin{equation}
x=a(\cos\xi-e), \hspace{0.4cm} y=a\sqrt{(1-e^2)}\sin\xi, \hspace{0.4cm} \Omega t=\xi-e\sin\xi,
\label{eq:16}
\end{equation}
where $a$ denotes the semi-major axis of the orbit with eccentricity $e$ and $\xi$ denotes the eccentric anomaly. As the angular velocity in an eccentric orbit varies with time, the Fourier transform of the current density can be expressed as a sum over the $n$-th harmonics of the fundamental orbital frequency $\Omega=\sqrt{G\frac{(m_1+m_2)}{a^3}}$. Following \cite{Poddar:2021yjd}, we calculate the position and velocity components of the binary in a Fourier space and hence, the stress-energy tensor components in Fourier space becomes
\begin{equation}
T_{xx}(\omega^\prime)=-\frac{\mu\omega^{\prime}{}^2}{4 n}\Big[J_{n-2}(ne)-2eJ_{n-1}(ne)+2eJ_{n+1}(ne)-J_{n+2}(ne)\Big],    
\label{eq:17}
\end{equation}
\begin{equation}
T_{yy}(\omega^\prime)=\frac{\mu\omega^{\prime2}a^2}{4n}\Big[J_{n-2}(ne)-2eJ_{n-1}(ne)+\frac{4}{n}J_n(ne)+2eJ_{n+1}(ne)-J_{n+2}(ne)\Big], 
\label{eq:18}
\end{equation}
\begin{equation}
T_{xy}(\omega^\prime)=-i\frac{\mu\omega^{\prime}{}^2 a^2 \sqrt{1-e^2}}{4n}\Big[J_{n+2}(ne)-2J_n(ne)+J_{n-2}(ne)\Big],
\label{eq:19}
\end{equation}
where $\omega^\prime=n\Omega$ and $J_n(x)$ denotes the Bessel function. Using Eqs. \ref{eq:17}, \ref{eq:18}, and \ref{eq:19}, we obtain two important relations
\begin{equation}
T_{ij}(\omega^{\prime})T^{ij*}(\omega^{\prime})=4\mu^2\omega'^4a^4\left(f(n,e)+\frac{J^2_n(ne)}{12n^4}\right), \hspace{0.5cm} \vert T^i{}_i\vert^2= \frac{\mu^2\omega^{\prime}{}^4a^4}{n^4}J^2_{n}(ne),    
\label{eq:20}
\end{equation}
where
\begin{equation}
\begin{split}
f(n,e)=\frac{1}{32n^2}\Big\{[J_{n-2}(ne)-2eJ_{n-1}(ne)+2eJ_{n+1}(ne)+\frac{2}{n}J_n(ne)-J_{n+2}(ne)]^2+\\
(1-e^2)[J_{n-2}(ne)-2J_n(ne)+J_{n+2}(ne)]^2+\frac{4}{3n^2}J^2_{n}(ne)\Big\}.
\end{split}
\label{eq:21}
\end{equation}
Therefore, the rate of energy loss due to the massless graviton radiation is obtained by using Eqs. \ref{eq:7}, \ref{eq:14} and \ref{eq:20} as
\begin{eqnarray}
\Big(\frac{dE}{dt}\Big)_A&=&\frac{1}{8(2\pi)^2}\frac{4}{M^2_{pl}}\int\frac{8\pi}{5}\Big[T_{ij}(\omega^\prime)T^*_{ji}(\omega^\prime)-\frac{1}{3}|T^{i}{}_{i}(\omega^\prime)|^2\Big]
\delta(\omega-\omega^\prime)\omega^2 d\omega,\nonumber\\
&=& \frac{32G}{5}\sum^\infty_{n=1}(n\Omega)^2\mu^2a^4(n\Omega)^4f(n,e)\nonumber\\
&=&\frac{32G}{5}\Omega^6\mu^2a^4(1-e^2)^{-7/2}\Big(1+\frac{73}{24}e^2+\frac{37}{96}e^4\Big).
\label{eq:22}
\end{eqnarray}
This is the well known Peters and Mathews result of GW radiation from two point masses in a Keplerian orbit \cite{Peters:1963ux}. This result coincides with the expression for massless graviton radiation derived in the context of standard (i.e. ghostly) and ghost-free fourth-order gravity theories, where the massless spin-$2$ mode is a propagating mode.

\subsection{Rate of energy loss $(dE/dt)_B$ due to massive graviton radiation}\label{subsecb}

The dispersion relation for the massive graviton mode is $|\textbf{k}|^2=\omega^2\Big(1-\frac{m^2_{2^+}}{\omega^2}\Big)$ and from the current conservation relation we obtain the stress tensor components as
\begin{equation}
T_{0j}=-\sqrt{1-\frac{m^2_{2^+}}{\omega^2}}\hat{k^i}T_{ij},\hspace{0.5cm} T_{00}=\Big(1-\frac{m^2_{2^+}}{\omega^2}\Big)\hat{k^i}\hat{k^j}T_{ij},
\label{eq:23}
\end{equation}
where the unit vector along the momentum direction of the massive graviton mode is $\hat{k^i}=\frac{k^i}{\omega\sqrt{1-\frac{m^2_2}{\omega^2}}}$. Therefore, the expression within the square bracket of Eq. \ref{eq:8} is rewritten as
\begin{equation}
|T_{\mu\nu}(k^\prime)|^2-\frac{1}{3}|T^{\mu}{}_{\mu}(k^\prime)|^2={\Lambda_{ij,lm}^B}T^{ij*}T^{lm},
\label{eq:24}
\end{equation}
where
\begin{equation}
\begin{split}
\Lambda_{ij,lm}^B= \Big[ \delta_{il}\delta_{jm}+\frac{2}{3}\Big(1-\frac{m^2_{2^+}}{\omega^2}\Big)^2\hat{k}_i\hat{k}_j\hat{k}_l\hat{k}_m -2\Big(1-\frac{m^2_{2^+}}{\omega^2}\Big) \hat{k}_j\hat{k}_m\delta_{il} -\frac{1}{3}\delta_{ij}\delta_{lm}+\frac{1}{3}\Big(1-\frac{m^2_{2^+}}{\omega^2}\Big)\times\\
(\delta_{ij}\hat{k}_l\hat{k}_m+\delta_{lm}\hat{k}_i\hat{k}_j) \Big].
\label{eq:25}
\end{split}
\end{equation}
The angular integration for the massive graviton mode is performed as
\begin{eqnarray}
\int d\Omega_k \Lambda_{ij,lm}^BT^{ij*}(\omega^\prime)T^{lm}({\omega^\prime})&=& \frac{8\pi}{5}\left(\left[\frac{5}{2}-\frac{5}{3}\left(1-\frac{m^2_{2^+}}{\omega'^2}\right)+\frac{2}{9}\left(1-\frac{m^2_{2^+}}{\omega'^2}\right)^2\right]T^{ij}T^*_{ij}\right.\nonumber\\
&& \left. +\left[-\frac{5}{6}+\frac{5}{9}\left(1-\frac{m^2_{2^+}}{\omega'^2}\right)+\frac{1}{9}\left(1-\frac{m^2_{2^+}}{\omega'^2}\right)^2\right]\vert T^i{}_{i}\vert^2\right).
\label{eq:26}
\end{eqnarray}
Thus, the rate of energy loss due to the massive graviton mode is
\begin{eqnarray}
\Big(\frac{dE}{dt}\Big)_B&=&  \frac{8G}{5}\Big(\frac{2m^2_{2^+}\alpha_2}{M^2_{pl}}\Big)\int \left[\left\lbrace\frac{5}{2}-\frac{5}{3}\left(1-\frac{m^2_{2^+}}{\omega'^2}\right)+\frac{2}{9}\left(1-\frac{m^2_{2^+}}{\omega'^2}\right)^2\right\rbrace T^{ij}T^*_{ij}\right.\nonumber\\
&& \left. +\left\lbrace -\frac{5}{6}+\frac{5}{9}\left(1-\frac{m^2_{2^+}}{\omega'^2}\right)+\frac{1}{9}\left(1-\frac{m^2_{2^+}}{\omega'^2}\right)^2\right\rbrace\vert T^i{}_{i}\vert^2\right]\delta(\omega-\omega^\prime)\omega^2 \Big(1-\frac{m^2_{2^+}}{\omega^2}\Big)^\frac{1}{2}d\omega. \nonumber\\
\label{eq:27}
\end{eqnarray}
Integrating Eq. \ref{eq:27} over all frequencies, we obtain 
\begin{equation}
\begin{split}
 \Big(\frac{dE}{dt}\Big)_B =\frac{32 G}{5}\Big(\frac{2m^2_{2^+}\alpha_2}{M^2_{pl}}\Big)\mu^2\Omega^6a^4 \sum_{n>n_{2+}}  n^6 \sqrt{1-\frac{n_{2+}^2}{n^2}} \Big[ \Big(\frac{19}{18}+\frac{11}{9}\frac{n_{2+}^2}{n^2}+\frac{2}{9}\frac{n_{2+}^4}{n^4}\Big)f(n,e)
+\\
\frac{5}{108}\Big(1-2\frac{n_{2+}^2}{n^2}+\frac{n_{2+}^4}{n^4}\Big)\frac{J^2_n(ne)}{n^4} \Big],
\label{eq:28}
\end{split}
\end{equation}
where $n_{2+}=m_{2+}/\Omega$ and $f(n,e)$ is given in Eq. \ref{eq:21}.
We can also write Eq. \ref{eq:28} to the leading order in $n_{2+}^2$ as
\begin{equation}
\begin{split}
\Big(\frac{dE}{dt}\Big)_B\simeq  \frac{32G}{5} \Big(\frac{2m^2_{2^+}\alpha_2}{M^2_{pl}}\Big)\mu^2 a^4\Omega^6 \Big[\sum_{n=1}^{\infty}\Big(\frac{19}{18}n^6f(n,e)+\frac{5}{108}n^2J^2_n(ne)\Big)+\\
n^2_{2+}\sum_{n=1}^{\infty}\Big(\frac{25}{36}n^4f(n,e)-
\frac{25}{216}J^2_n(ne)\Big)\Big]
+\mathcal{O}(n^4_{2+}).
\label{eq:29}
\end{split}
\end{equation} 
Therefore, the energy loss rate associated with massive graviton radiation is further suppressed by a factor of $\mathcal{O}(G)$ relative to that of the massless graviton mode.

\subsection{Rate of energy loss $(dE/dt)_C$ due to massive scalar radiation}\label{subsecc}

The dispersion relation for the massive scalar mode is $|\textbf{k}|^2=\omega^2\Big(1-\frac{m^2_{0+}}{\omega^2}\Big)$ and the current conservation relation yields
\begin{equation}
T_{0j}=-\sqrt{1-\frac{m^2_{0+}}{\omega^2}}\hat{k^i}T_{ij},\hspace{0.5cm} T_{00}=\Big(1-\frac{m^2_{0+}}{\omega^2}\Big)\hat{k^i}\hat{k^j}T_{ij},
\label{eq:31}
\end{equation}

where $\hat{k^i}=\frac{k^i}{\omega\sqrt{1-\frac{m^2_{0+}}{\omega^2}}}$.

Thus, the term within the square bracket of Eq. \ref{eq:9} can be written as
\begin{equation}
\Big[\frac{1}{3}|T^{\mu}{}_{\mu}(k^\prime)|^2\Big] = {\Lambda_{ij,lm}^C}T^{ij*}T^{lm},
\label{eq:32}
\end{equation}
where 
\begin{equation}
\Lambda_{ij,lm}^C= \frac{1}{3} \Big[ \delta_{ij}\delta_{lm} + {\Big(1-\frac{m^2_{0+}}{\omega^2}\Big)}^2 \hat{k_i}\hat{k_j}\hat{k_l}\hat{k_m} - \Big(1-\frac{m^2_{0+}}{\omega^2}\Big) (\delta_{ij}\hat{k_l}\hat{k_m} + \delta_{lm}\hat{k_i}\hat{k_j} ) \Big].
\label{eq:33}
\end{equation}
Therefore, we can perform the  angular integral as 
\begin{equation}
\begin{split}
 \int d\Omega_k \Lambda_{ij,lm}^CT^{ij*}(\omega^\prime)T^{lm}({\omega^\prime}) = \frac{8\pi}{5} \Big[ \Big \{ \frac{1}{9}{\Big(1-\frac{m^2_{0+}}{\omega^2}\Big)}^2 \Big\} T^{ij*}(\omega^\prime)T^{ij}({\omega^\prime})+ \Big \{ \frac{5}{6}+\\
 \frac{1}{18}{\Big(1-\frac{m_{0+}^2}{\omega^2}\Big)}^2-\frac{5}{9}{\Big(1-\frac{m^2_{0+}}{\omega^2}\Big)} \Big\}|T^{i}{}_{i}(\omega^\prime)|^2\Big].
 \end{split}
 \label{eq:34}
\end{equation}
From Eq. \ref{eq:9}, we obtain 
\begin{equation}
\begin{split}
\Big(\frac{dE}{dt}\Big)_C=\frac{1}{8(2\pi)^2}\Big(\frac{8 m_{0^+}^2 (3\alpha_1+\alpha_2)}{M_{Pl}^4}\Big)\frac{8\pi}{5} \int \Big[ \Big \{ \frac{1}{9}{\Big(1-\frac{m^2_{0+}}{\omega^2}\Big)}^2 \Big\} T^{ij*}(\omega^\prime)T^{ij}({\omega^\prime})\\
+ \Big \{ \frac{5}{6}+\frac{1}{18}{\Big(1-\frac{m^2_{0+}}{\omega^2}\Big)}^2- \frac{5}{9}{\Big(1-\frac{m^2_{0+}}{\omega^2}\Big)} \Big\}|T^{i}{}_{i}(\omega^\prime)|^2\Big]\delta(\omega-\omega^\prime)\omega^2 \Big(1-\frac{m^2_{0+}}{\omega^2}\Big)^\frac{1}{2}d\omega.
\end{split}
\label{eq:35}
\end{equation}
Evaluating Eq. \ref{eq:35} for a binary orbit in the $x-y$ plane and $\omega^\prime=n\Omega$ as done in the previous cases, one infers
\begin{equation}
\begin{split}
\Big(\frac{dE}{dt}\Big)_C =\frac{1}{8(2\pi)^2}\Big(\frac{8 m_{0^+}^2 (3\alpha_1+\alpha_2)}{M_{Pl}^4}\Big)\frac{8\pi}{5} \sum_{n>n_{0+}} \Big[ \Big \{ \frac{1}{9} \Big(1-2\frac{n_{0+}^2}{n^2}+\frac{n_{0+}^4}{n^4}\Big) \Big\} 4\mu^2n^4\Omega^4a^4 \Big(f(n,e)+\\
\frac{J^2_n(ne)}{12n^4}\Big)+ \Big \{ \frac{5}{6}+\frac{1}{18}\Big(1-2\frac{n_{0+}^2}{n^2}+\frac{n_{0+}^4}{n^4}\Big)-\frac{5}{9}\Big(1-\frac{n_{0+}^2}{n^2}\Big) \Big\}\mu^2\Omega^4a^4 J^2_{n}(ne)\Big]n^2\Omega^2 \Big(1-\frac{n_{0+}^2}{n^2}\Big)^\frac{1}{2},
\end{split}
\label{eq:36}
\end{equation}
where $n_{0+}=m_{0+}/\Omega$. We write Eq. \ref{eq:36} in a more compact form as
\begin{equation}
\begin{split}
\Big(\frac{dE}{dt}\Big)_C=\frac{32G}{5}\Big(\frac{4m^2_{0+}(3\alpha_1+\alpha_2)}{M_{pl}^2}\Big)\mu^2 a^4\Omega^6\sum_{n>n_{0+}} n^6\sqrt{1-\frac{n^2_{0+}}{n^2}}\Big[\frac{1}{18}\Big(1-\frac{2n^2_{0+}}{n^2}+\frac{n^4_{0+}}{n^4}\Big)f(n,e)+\\
\frac{5}{108}\Big(1+\frac{n^2_{0+}}{n^2}+\frac{1}{4}\frac{n^4_{0+}}{n^4}\Big)\frac{J^2_n(ne)}{n^4}\Big].    
\end{split}
\label{eq:37}
\end{equation}
We can also expand Eq. \ref{eq:37} up to $\mathcal{O}(n^2_{0+})$ as
\begin{equation}
\begin{split}
\Big(\frac{dE}{dt}\Big)_C=\frac{32G}{5}\Big(\frac{4m^2_{0+}(3\alpha_1+\alpha_2)}{M^2_{pl}}\Big)\mu^2 a^4\Omega^6\Big[\sum_{n=1}^\infty\Big(\frac{1}{18}n^6f(n,e)+\frac{5}{108}n^2J^2_n(ne)\Big)-\\
n^2_{0+}\sum_{n=1}^\infty\Big(\frac{5}{36}n^4f(n,e)-\frac{5}{216}J^2_n(ne)\Big)\Big]+\mathcal{O}(n^4_{0+}).
\end{split}
\label{eq:38}
\end{equation}
The energy loss due to massive scalar radiation is suppressed by an additional factor of $\mathcal{O}(G)$ relative to the massless graviton contribution. Both massive graviton and massive scalar radiation exhibit the same overall suppression of $\mathcal{O}(G^2)$ in their respective energy loss rates.

\subsection{Total energy-loss rate}

The total energy-loss rate in ghost-free fourth-order gravity is obtained by combining Eqs. \ref{eq:22}, \ref{eq:28}, and \ref{eq:37} as

\begin{equation}
\begin{split}
\Big(\frac{dE}{dt}\Big)_{A+B+C}=\frac{32G}{5}\mu^2a^4\Omega^6\Bigg[f(e)+\Big(\frac{2m^2_{2+}\alpha_2}{M_{pl}^2}\Big)
\sum_{n>n_{2+}}  n^6 \sqrt{1-\frac{n_{2+}^2}{n^2}}
\Big[ \Big(\frac{19}{18}+\frac{11}{9}\frac{n_{2+}^2}{n^2}+\frac{2}{9}\frac{n_{2+}^4}{n^4}\Big)\times\\
 f(n,e)
+
\frac{5}{108}\Big(1-2\frac{n_{2+}^2}{n^2}+\frac{n_{2+}^4}{n^4}\Big)\frac{J^2_n(ne)}{n^4}\Big]+
\Big(\frac{4m^2_{0+}(3\alpha_1+\alpha_2)}{M_{pl}^2}\Big)\sum_{n>n_{0+}} n^6\sqrt{1-\frac{n^2_{0+}}{n^2}}
\times\\
\Big[\frac{1}{18}
\Big(1-\frac{2n^2_{0+}}{n^2}+\frac{n^4_{0+}}{n^4}\Big)f(n,e)+
\frac{5}{108}\Big(1+\frac{n^2_{0+}}{n^2}+\frac{1}{4}\frac{n^4_{0+}}{n^4}\Big)\frac{J^2_n(ne)}{n^4}\Big]\Bigg],
\end{split}
\end{equation}
where 
\[
f(e)=(1-e^2)^{-7/2}[1+(73/24)e^2+(37/96)e^4],
\]
and $f(n,e)$ is defined in Eq. \ref{eq:21}. The energy loss rate associated with all three propagating modes-massless spin-$2$, massive spin-$2$, and massive spin-$0$ is proportional to $\mu^2 a^4 \Omega^6$. Radiation of the massive spin-2 and scalar modes is kinematically allowed only when their respective masses satisfy $m_{2^+}, m_{0^+} < \Omega$ for the fundamental harmonic mode ($n = 1$). This imposes direct limits of validity on $m_{2^+}$ and $m_{0^+}$, as $\Omega$ is a measurable quantity for a binary system.

 In the limit $m_{2^+}, m_{0^+} \rightarrow 0$ of ghost-free quadratic gravity theory, radiation from the massive modes vanishes, and only the massless graviton contributes to energy loss. In the limit $\alpha_1 \to 0$, both massive modes contribute to the radiation, whereas for $\alpha_2 \to 0$, only the massive scalar mode contributes alongside the massless graviton.

\subsection{Radiation formulae in standard fourth-order gravity}
\label{subsec:radiation_standard4thordergravity}

In order to compare with the usual fourth-order gravity theory, we chose the limit for large torsion mass parameter $M_T\gg M_{pl}$ and hence $m^2_{2+}\rightarrow -M^2_{pl}/2\alpha_2$ Eq. \ref{eq:28} in usual fourth-order gravity becomes
\begin{equation}
\begin{split}
 \Big(\frac{dE}{dt}\Big)_{B^\prime} =-\frac{32 G}{5}\mu^2\Omega^6a^4 \sum_{n>n_{2+}}  n^6 \sqrt{1-\frac{n_{2+}^2}{n^2}} \Big[ \Big(\frac{19}{18}+\frac{11}{9}\frac{n_{2+}^2}{n^2}+\frac{2}{9}\frac{n_{2+}^4}{n^4}\Big)f(n,e)
+\\
\frac{5}{108}\Big(1-2\frac{n_{2+}^2}{n^2}+\frac{n_{2+}^4}{n^4}\Big)\frac{J^2_n(ne)}{n^4} \Big],
\label{eq:30}
\end{split}
\end{equation}
where $n_{2+}=m_{2+}/\Omega$ within the summation sign and the additional negative sign on the right-hand side reflects that the massive graviton behaves as a ghost mode. To the leading order in $n^2_{2+}$, Eq. \ref{eq:30} becomes
\begin{equation}
\begin{split}
\Big(\frac{dE}{dt}\Big)_{B^\prime}\simeq - \frac{32G}{5} \mu^2 a^4\Omega^6 \Big[\sum_{n=1}^{\infty}\Big(\frac{19}{18}n^6f(n,e)+\frac{5}{108}n^2J^2_n(ne)\Big)+n^2_{2+}\sum_{n=1}^{\infty}\Big(\frac{25}{36}n^4f(n,e)-\\
\frac{25}{216}J^2_n(ne)\Big)\Big]+\mathcal{O}(n^4_{2+}).  
\end{split}
\end{equation}

In the limit of infinitely heavy torsion mass parameter, $m^2_{0+}\rightarrow M^2_{pl}/4(3\alpha_1+\alpha_2)$ and Eq. \ref{eq:37} becomes 
\begin{equation}
\begin{split}
\Big(\frac{dE}{dt}\Big)_{C^\prime}=\frac{32G}{5}\mu^2 a^4\Omega^6\sum_{n>n_{0+}} n^6\sqrt{1-\frac{n^2_{0+}}{n^2}}\Big[\frac{1}{18}\Big(1-\frac{2n^2_{0+}}{n^2}+\frac{n^4_{0+}}{n^4}\Big)f(n,e)+\\
\frac{5}{108}\Big(1+\frac{n^2_{0+}}{n^2}+\frac{1}{4}\frac{n^4_{0+}}{n^4}\Big)\frac{J^2_n(ne)}{n^4}\Big],   
\end{split}
\label{eq:39}
\end{equation}
which is the rate of energy loss expression for massive scalar emission in the standard fourth-order gravity theories. To the leading order in $n^2_{0+}$, Eq. \ref{eq:39} becomes
\begin{equation}
\begin{split}
 \Big(\frac{dE}{dt}\Big)_{C^\prime} \simeq \frac{32 G}{5}\mu^2\Omega^6a^4 \Big[ \sum_{n=1}^\infty \Big( \frac{1}{18} n^6 f(n,e) + \frac{5}{108} n^2 J_n^2(ne) \Big) -n_{0+}^2\Big( \frac{5}{36}n^4 f(n,e) -\\
 \frac{5}{216}J_n^2(ne)\Big) \Big] +\mathcal{O}(n_{0+}^4).    
 \end{split}
 \label{eq:40}
\end{equation}

For conventional fourth-order gravity theories, the vanishing mass limit $m_{2^+}, m_{0^+} \rightarrow 0$ leads to a complete suppression of radiation at the leading order, i.e, the quadrupolar radiation formula is not recovered (sum of Eqs. \ref{eq:22}, \ref{eq:30}, \ref{eq:39} results zero at the order of $\Omega^6$). 
In the standard fourth-order gravity theory which has a ghost, the dominant contribution to the energy loss rate arises at order $\mathcal{O}(m_{2^+}^2, m_{0^+}^2)$ and $\Omega^4$ and hence, the total energy loss rate becomes 
\begin{equation}
\begin{split}
\Big(\frac{dE}{dt}\Big)_{A+B^\prime+C^\prime}=\frac{32G}{5}\mu^2 a^4\Omega^4 m^2_{2+}\sum_{n=1}^\infty\Big(\frac{25}{216}J^2_n(ne)-\frac{25}{36}n^4 f(n,e)\Big)+\\
\frac{32G}{5}\mu^2 a^4\Omega^4 m^2_{0+}\sum_{n=1}^\infty \Big(\frac{5}{216}J^2_n(ne)-\frac{5}{36}n^4 f(n,e)\Big).   
\end{split}
\label{eq:41}
\end{equation}
Therefore, to obtain the gravitational quadrupole radiation at $\Omega^6$ order within the framework of standard fourth-order gravity, we can set the massive spin-2 ghost infinitely heavy, decoupling it from the low-energy dynamics. As a result, the gravitational sector of the low-energy effective theory contains only the massless spin-2 mode and a massive spin-0 scalar mode. Thus, the total rate of energy loss in standard fourth-order gravity theory (the ghost is integrated out) is
\begin{equation}
\begin{split}
\Big(\frac{dE}{dt}\Big)_{A+C^\prime}=\frac{32G}{5}\mu^2a^4\Omega^6(1-e^2)^{-7/2}\Big(1+\frac{73}{24}e^2+\frac{37}{96}e^4\Big)+\frac{32G}{5}\mu^2 a^4\Omega^6\times\\
\sum_{n>n_{0+}} n^6\sqrt{1-\frac{n^2_{0+}}{n^2}}
\Big[\frac{1}{18}\Big(1-\frac{2n^2_{0+}}{n^2}+\frac{n^4_{0+}}{n^4}\Big)f(n,e)+
\frac{5}{108}\Big(1+\frac{n^2_{0+}}{n^2}+\frac{1}{4}\frac{n^4_{0+}}{n^4}\Big)\frac{J^2_n(ne)}{n^4}\Big].
\end{split}
\label{eq:42}
\end{equation}
It is worth noting that, even in the absence of ghost $(m_{2+})$, the leading-order energy loss rate is enhanced by an additional factor of $1/18$. The rate of orbital period decay is directly linked to the energy loss rate. In the following, we use observational data from the orbital period decay of various binary systems to constrain the masses and couplings of the different propagating modes.

\section{Orbital period decay for a quasi-stable orbit}\label{sec5}

In ghost-free quadratic gravity, the gravitational potential is modified relative to the standard Newtonian potential, resulting in a new force law between two compact objects in a binary system. The modifications to the potential arise from both the massive spin-2 graviton and massive spin-0 scalar modes, each contributing a Yukawa-type correction. 

In the ghost-free quadratic gravity framework, the gravitational force between two stars in a quasi-stable binary orbit is obtained by differentiating the modified gravitational potential given in Eq. \ref{pot6}, yielding
\begin{equation}
|\mathbf{F}| = \frac{G m_1 m_2}{a^2} \left[1 + \alpha e^{-m_{2+} a}(1 + m_{2+} a) + \beta e^{-m_{0+} a}(1 + m_{0+} a)\right],
\label{n1}
\end{equation}
where the correction coefficients are defined as
\begin{equation}
\alpha = \frac{8}{3} \left(\frac{m^2_{2+} \alpha_2}{M^2_{\rm pl}}\right), \qquad \beta = \frac{4}{3} \frac{m^2_{0+}(3\alpha_1 + \alpha_2)}{M^2_{\rm pl}},
\label{n2}
\end{equation}
where, $a$ denotes the semi-major axis of the binary orbit, assumed to remain constant over time. The corresponding orbital frequency is then given by
\begin{equation}
\Omega^2 = \frac{G(m_1 + m_2)}{a^3} \left[1 + \alpha e^{-m_{2+} a}(1 + m_{2+} a) + \beta e^{-m_{0+} a}(1 + m_{0+} a)\right].
\label{n3}
\end{equation}

The total energy of the binary system in this framework becomes
\begin{equation}
E_{\rm tot} = -\frac{G m_1 m_2}{2 a} \left[1 + \alpha e^{-m_{2+} a}(1 + m_{2+} a) + \beta e^{-m_{0+} a}(1 + m_{0+} a)\right].
\label{n4}
\end{equation}

Since the orbital period $P$ is related to the orbital frequency $\Omega$ via $P = 2\pi/\Omega$, the rate of orbital period decay can be connected to the rate of energy loss as $\dot{P} = (dP/da)(da/dE)(dE/dt)$.

Using Eqs. \ref{n3} and \ref{n4}, we obtain the expression for the rate of orbital period decay in the ghost-free fourth-order gravity theory as
\begin{equation}
\begin{split}
\dot{P} = 6\pi a^{5/2} G^{-3/2} (m_1 + m_2)^{-1/2} (m_1 m_2)^{-1} \left(\frac{dE}{dt}\right)_{A+B+C} \times \Bigg(1 - \frac{3}{2} \alpha e^{-m_{2+} a}(1 + m_{2+} a) \\
-\frac{2}{3} \alpha m^2_{2+} a^2 e^{-m_{2+} a} - \frac{3}{2} \beta e^{-m_{0+} a}(1 + m_{0+} a) - \frac{2}{3} \beta m^2_{0+} a^2 e^{-m_{0+} a} \Bigg).
  \end{split}
  \label{n5}
  \end{equation}

The observed orbital period decay $\dot{P}$, measured through pulsar timing, provides direct access to the rate of energy loss $dE/dt$. Given that theoretical expressions for $dE/dt$ are derived in the ghost-free fourth-order gravity framework, these observations can be used to place constraints on the coupling strengths and mass parameters associated with these effective theories of gravity.

\vspace{1.0cm}\noindent {\bf Limit for standard fourth-order gravity:} In the standard fourth-order gravity theory, $\alpha\rightarrow (-4/3)$ and $\beta\rightarrow(1/3)$ and we obtain the rate of orbital period loss as
\begin{equation}
\begin{split}
\dot{P} = 6\pi a^{5/2} G^{-3/2} (m_1 + m_2)^{-1/2} (m_1 m_2)^{-1} \left(\frac{dE}{dt}\right)_{A+B^\prime+C^\prime} \times \Bigg(1 +2 e^{-m_{2+} a}(1 + m_{2+} a) \\
+\frac{8}{9} m^2_{2+} a^2 e^{-m_{2+} a} - \frac{1}{2}  e^{-m_{0+} a}(1 + m_{0+} a) - \frac{2}{9}  m^2_{0+} a^2 e^{-m_{0+} a} \Bigg).
  \end{split}
  \label{n5a}
\end{equation}
Using this formula, the observed orbital period decay $\dot{P}$ can place constraints on the standard fourth-order gravity theory.

\section{Radiation of massless and massive modes from coalescence binaries}\label{sec6}

In the case of a coalescing binary, the orbital separation and consequently the orbital frequency evolves over time. Although the system may initially exhibit eccentricity, it tends to circularize as it evolves before being detected by the GW detectors. Therefore, in this section, we focus on binaries with negligible eccentricity.

The instantaneous gravitational force in the ghost-free fourth-order gravity theory is obtained by differentiating the modified potential of Eq. \ref{pot6}. This yields the force law in Eq. \ref{n1}, evaluated at the instantaneous separation by substituting $a \to r(t)$ for the coalescing binary. Likewise, the instantaneous orbital frequency follows from Eq. \ref{n3} with the same substitution $a \to r(t)$.

We focus on the GW170817 event detected by LIGO to constrain the coupling and mass parameters of ghostful and ghost-free fourth-order gravity theories. Since the LIGO sensitivity band begins at a GW frequency $f_{\mathrm{GW}} = \Omega/\pi\sim \mathcal{O}(10~\mathrm{Hz})$ \cite{Abbott:2016xvh}, a binary system of NSs, each with mass $1.25~M_\odot$, enters this band when their orbital separation is approximately $\mathcal{O}(700~\mathrm{km})$, as estimated from Eq. \ref{n3} under the assumption that standard Newtonian gravity dominates at that scale.

Given a typical NS radius of $\sim 10~\mathrm{km}$, the range of the Yukawa-type corrections due to the massive spin-2 and spin-0 modes that LIGO could probe lies within $m^{-1}=\{m^{-1}_{2+}, m^{-1}_{0+}\} \sim \mathcal{O}(20~\mathrm{km} - 750~\mathrm{km})$ which corresponds to masses $m=\{m_{2+}, m_{0+}\} \sim \mathcal{O}(10^{-11}~\mathrm{eV} - 3 \times 10^{-13}~\mathrm{eV})$.

The effects of these additional modes could be observable if the range of the new force ($m^{-1}$) exceeds the orbital separation. Conversely, if the binary separation is larger than the range of the new interaction, LIGO would not be sensitive to these modes. Moreover, in the case of an infinitely long-range force, the Yukawa corrections become indistinguishable from the standard Newtonian potential for the ghost-free fourth order gravity theory.

We can write the total instantaneous energy of the coalescing binary system as Eq. \ref{n4} with the substitution $a\to r(t)$. Hence, the rate of total energy loss is obtained as
\begin{equation}
\frac{dE_{tot}}{dt}=\frac{Gm_1m_2}{2r^2}\frac{dr}{dt}\Big[1+\alpha e^{-m_{2+}r}(1+m_{2+}r+m^2_{2+}r^2)+\beta e^{-m_{0+}r}(1+m_{0+}r+m^2_{0+}r^2)\Big].    
\label{rad5}
\end{equation}
In the following, we consider the limit where the orbital eccentricity vanishes $e \rightarrow 0$ because, by the time the coalescing binary enters the LIGO/Virgo frequency band, the orbit has effectively circularized due to GW emission. 

Therefore, in the vanishing eccentricity limit, we obtain the rate of energy loss terms for the ghost-free fourth-order gravity theory as
\begin{equation}
\begin{aligned}
\Big(\frac{dE}{dt}\Big)_A&=\frac{32G}{5}\mu^2r^4\Omega^6,\\
\Big(\frac{dE}{dt}\Big)_B&=\frac{32G}{5}\Big(\frac{2m^2_{2+}\alpha_2}{M^2_{pl}}\Big)\mu^2 r^4 \Omega^6\sqrt{1-\frac{m^2_{2+}}{4\Omega^2}}\Big[\frac{19}{18}+\frac{11}{36}\Big(\frac{m_{2+}}{\Omega}\Big)^2+\frac{1}{72}\Big(\frac{m_{2+}}{\Omega}\Big)^4\Big],\\
\Big(\frac{dE}{dt}\Big)_C&=\frac{32G}{5}\Big(\frac{4m^2_{0+}(3\alpha_1+\alpha_2)}{M^2_{pl}}\Big)\mu^2 r^4\Omega^6\times \frac{1}{18}\Big(1-\frac{m^2_{0+}}{4\Omega^2}\Big)^{5/2}.
\end{aligned}
\label{kl1}
\end{equation}

If the energy loss arises solely from the radiation of the massless spin-2 graviton mode, then the only deviation from GR stems from the modification of the gravitational force law as shown in Eqs. \ref{n5} and \ref{n5a}. This regime corresponds to the case when the mass of the massive modes is greater than $2\Omega$ but less than $1/a$ and when the modes behave as mediator. In this case, only the modified force contributes to the orbital period decay, while radiation from off-shell massive modes is absent.

In the following, we evaluate the rate of change of the orbital frequency for the GW170817 event within the ghost-free fourth-order gravity, expressing the time-dependent separation in terms of $\Omega$. (For the corresponding quantity in the standard, i.e. ghostly, fourth-order gravity theory, see the last paragraph of this section.) This formulation is important as the variation of orbital frequency with time is measurable from LIGO/Virgo. By equating the total energy-loss rate, $(dE/dt)_{\text{tot}}$ from Eq.~~\ref{rad5}, with the energy-loss rate due to the radiation of massless, massive graviton and massive spin-$0$ scalar mode, $-((dE/dt)_A+(dE/dt)_B+(dE/dt)_C)$ from Eq. \ref{kl1}, we obtain 

\begin{equation}
\begin{split}
\Big(\frac{dr}{dt}\Big)=-\frac{64}{5}\frac{G^3m_1m_2(m_1+m_2)}{r^3}\Big[1+2\alpha e^{-m_{2+}r}(1+m_{2+}r)-\alpha m^2_{2+}r^2 e^{-m_{2+}r}+\\
2\beta e^{-m_{0+}r}(1+m_{0+}r)-\beta m^2_{0+}r^2e^{-m_{0+}r}\Big]\Big[1+\frac{2m^2_{2+}\alpha_2}{M_{pl}^2}\sqrt{1-\frac{m^2_{2+}}{4\Omega^2}}\Big(\frac{19}{18}+\\
\frac{11}{36}\frac{m^2_{2+}}{\Omega^2}+\frac{1}{72}\frac{m^4_{2+}}{\Omega^4}\Big)+\frac{4m^2_{0+}(3\alpha_1+\alpha_2)}{M^2_{pl}}\frac{1}{18}\Big(1-\frac{m^2_{0+}}{4\Omega^2}\Big)^{5/2}\Big],
\end{split}
\label{rad6}
\end{equation}
where, in the limit $m_{2+},m_{0+}\rightarrow 0$, Eq. \ref{rad6} reduces to the GR result
\begin{equation}
\frac{dr}{dt}=-\frac{64}{5}\frac{G^3m_1m_2(m_1+m_2)}{r^3}.
\label{rad7}
\end{equation}
Differentiating $\Omega$ (Eq. \ref{n3} with $a\to r(t)$) with respect to time and using Eq. \ref{rad6}, we obtain
\begin{equation}
\begin{split}
\Omega\dot{\Omega}=\frac{96}{5}\frac{G^4(m_1+m_2)^2m_1m_2}{r^7}\Big[1+3\alpha(1+m_{2+}r)e^{-m_{2+}r}-\frac{2}{3}\alpha m^2_{2+}r^2 e^{-m_{2+}r}+\\
3\beta(1+m_{0+}r)e^{-m_{0+}r}-\frac{2}{3}\beta m^2_{0+}r^2 e^{-m_{0+}r}\Big]\Big[1+\frac{2m^2_{2+}\alpha_2}{M_{pl}^2}\sqrt{1-\frac{m^2_{2+}}{4\Omega^2}}\Big(\frac{19}{18}+\\
\frac{11}{36}\frac{m^2_{2+}}{\Omega^2}+\frac{1}{72}\frac{m^4_{2+}}{\Omega^4}\Big)+\frac{4m^2_{0+}(3\alpha_1+\alpha_2)}{M^2_{pl}}\frac{1}{18}\Big(1-\frac{m^2_{0+}}{4\Omega^2}\Big)^{5/2}\Big].   
\end{split}
\label{rad8}
\end{equation}
Here, $r$ is a function of time, which can also be expressed in terms of the angular frequency by using the inversion of Eq. \ref{n3} with $a\to r$ as
\begin{equation}
\begin{split}
r(\Omega)=\Big[\frac{G(m_1+m_2)}{\Omega^2}\Big]^{1/3}\Big[1+\frac{\alpha}{3}\Big\{1+(G(m_1+m_2))^{1/3}\frac{m_{2+}}{\Omega^{2/3}}\Big\}\exp\Big(-\frac{(G(m_1+m_2))^{1/3}m_{2+}}{\Omega^{2/3}}\Big)+\\
\frac{\beta}{3}\Big\{1+(G(m_1+m_2))^{1/3}\frac{m_{0+}}{\Omega^{2/3}}\Big\}\exp\Big(-\frac{(G(m_1+m_2))^{1/3}m_{0+}}{\Omega^{2/3}}\Big)\Big].
\end{split}
\label{rad9}
\end{equation}
Using Eq. \ref{rad9}, we can write the rate of change of the angular frequency as
\begin{equation}
\begin{split}
\dot{\Omega}=\frac{96}{5}(G\mathcal{M}_{ch})^{5/3}\Omega^{11/3}\Big[1+\frac{2\alpha}{3}e^{-m_{2+}(\mathcal{M}\Omega^{-2})^{1/3}}\{1+m_{2+}(\mathcal{M}\Omega^{-2})^{1/3}-m^2_{2+}(\mathcal{M}\Omega^{-2})^{2/3}\}+\\
\frac{2\beta}{3}e^{-m_{0+}(\mathcal{M}\Omega^{-2})^{1/3}}\{1+m_{0+}(\mathcal{M}\Omega^{-2})^{1/3}-m^2_{0+}(\mathcal{M}\Omega^{-2})^{2/3}\}\Big]\times \Big[1+\frac{2m^2_{2+}\alpha_2}{M_{pl}^2}\sqrt{1-\frac{m^2_{2+}}{4\Omega^2}}\Big(\frac{19}{18}+\\
\frac{11}{36}\frac{m^2_{2+}}{\Omega^2}+\frac{1}{72}\frac{m^4_{2+}}{\Omega^4}\Big)+\frac{4m^2_{0+}(3\alpha_1+\alpha_2)}{M^2_{pl}}\frac{1}{18}\Big(1-\frac{m^2_{0+}}{4\Omega^2}\Big)^{5/2}\Big].    
\end{split}
\label{rad10}
\end{equation}
where we define $\mathcal{M}=G(m_1+m_2)$ and $\mathcal{M}_{ch}=(m_1m_2)^{3/5}/(m_1+m_2)^{1/5}$ is called the Chirp mass of a coalescing binary which is a measurable quantity. The rate of change of angular frequency of Eq. \ref{rad10} contains contributions from the massless and massive spin-$2$ modes and massive spin-$0$ scalar mode.

In the limit, $m_{2+}, m_{0+}\rightarrow 0$, $\alpha, \beta\rightarrow 0$, the total rate of change of angular frequency for the ghost-free gravity theory becomes
\begin{equation}
\dot{\Omega}=\frac{96}{5}(G\mathcal{M}_{ch})^{5/3}\Omega^{11/3},
\label{rad11}
\end{equation}
which is the standard formula for the rate of change of orbital frequency in GR. In solving the first-order differential equations (Eq. \ref{rad10}), we impose the boundary condition $f(0)=10~\mathrm{Hz}$, corresponding to the lower sensitivity threshold of LIGO. In addition, the reconstruction of the chirp mass $\mathcal{M}_{ch}=1.188^{+0.004}_{-0.002}M_{\odot}$, carries a larger uncertainty of about $0.4\%$ for the GW170817 event \cite{LIGOScientific:2017vwq}, arising from the unknown source distance. Although the Chirp mass has a much smaller uncertainty in the detector frame, as $\lesssim 0.067\%$. However, as a conservative choice, we choose the larger uncertainty to be $\sim 0.4\%$ in obtaining the bounds on the couplings and masses.

Therefore, for non-zero values of $m_{2+}$ and $m_{0+}$, observable deviations may arise in the LIGO/Virgo timing signal. Since these massive modes are only radiated when their masses are below the threshold $2\Omega$, this places an upper bound on their allowed values. The presence of such modes can lead to modifications in the inspiral dynamics, thereby affecting the inferred value of the chirp mass-an observable quantity extracted from the GW signal.

\textbf{Limit for standard fourth-order gravity:} In the standard fourth-order gravity theory, Eq. \ref{kl1} reduces to
\begin{equation}
\begin{aligned}
\Big(\frac{dE}{dt}\Big)_A&=\frac{32G}{5}\mu^2r^4\Omega^6,\\
\Big(\frac{dE}{dt}\Big)_{B^\prime}&=-\frac{32G}{5}\mu^2 r^4 \Omega^6\sqrt{1-\frac{m^2_{2+}}{4\Omega^2}}\Big[\frac{19}{18}+\frac{11}{36}\Big(\frac{m_{2+}}{\Omega}\Big)^2+\frac{1}{72}\Big(\frac{m_{2+}}{\Omega}\Big)^4\Big],\\
\Big(\frac{dE}{dt}\Big)_{C^\prime}&=\frac{32G}{5}\mu^2 r^4\Omega^6\times \frac{1}{18}\Big(1-\frac{m^2_{0+}}{4\Omega^2}\Big)^{5/2}.
\end{aligned}
\label{kl2}
\end{equation}
Therefore, the emission of massive modes in both cases is allowed for circular orbits provided that $m=\{m_{2+}, m_{0+}\} < 2\Omega$, obtained from the kinematic factors of Eqs. \ref{kl1} and \ref{kl2}.

The rate of angular frequency is obtained by substituting $\frac{2m^2_{2+}\alpha_2}{M^2_{\rm pl}} = -1$ to (\ref{rad10}) and by setting $\frac{4(3\alpha_1 + \alpha_2)m^2_{0+}}{M^2_{\rm pl}} = 1$ along with $\alpha = -\frac{4}{3}$ and $\beta = \frac{1}{3}$.

In the standard fourth-order gravity theory, radiation is entirely suppressed in the massless limit of the modes, resulting in a constant angular frequency over time ($d\Omega/dt\rightarrow 0$, at the leading order). However, when the modes are massive, the energy loss rate $\dot{E}$ scales as $\Omega^4$, and the rate of change of the angular frequency $\dot{\Omega}$ scales as $\Omega^{5/3}$. This leads to a discrepancy with respect to the GR result (Eq. \ref{rad11}). In contrast, the ghost-free fourth-order gravity theory remains free from this issue, even in the massive limit, with $\dot{\Omega}$ scaling as $\Omega^{11/3}$ at the leading order.

\section{Constraints from orbital period loss of binary systems}\label{sec7}

In the following, we use two quasi-stable binary systems, PSR B1913+16 \cite{Weisberg:2016jye} and PSR J1738+0333~\cite{Freire:2012mg} to derive constraints on the couplings and masses of the modes in standard and ghost-free fourth-order gravity. The Hulse-Taylor binary (PSR B1913+16) consists of a pulsar and a NS companion, while PSR J1738+0333 is composed of a pulsar and a WD companion. The orbital parameters of these systems, together with the observed orbital period decay and the corresponding GR predictions, are summarized in TABLE \ref{tableI}. In this context, the orbital period decay can receive contributions from the modified force law, from additional radiation channels, or from their combined effects, with each mechanism being relevant at specific mass ranges of the modes.

\begin{table}[h]
\centering
\resizebox{0.8\textwidth}{!}{
\begin{tabular}{ |l|c|c|c|c|c| }
 
 \hline
\textbf{Parameters} \hspace{0.01cm} & \textbf{PSR B1913+16}\hspace{0.01cm}& \textbf{PSR J1738+0333}\hspace{0.01cm}\\
 \hline
Pulsar mass $m_1$ (solar masses) &$1.438\pm 0.001$ &$1.46^{+0.06}_{-0.05}$ \\
Companion mass $m_2$ (solar masses)&$1.390\pm 0.001$ & $0.181^{+0.008}_{-0.007}$\\
Eccentricity $e$ &$0.6171340(4)$ &$(3.4\pm 1.1)\times10^{-7}$  \\
Orbital period $P$ (d)&$0.322997448918(3)$&$0.3547907398724(13)$\\
Intrinsic $\dot{P}(10^{-12}~\rm{s~s^{-1}})$ &$-2.398\pm 0.004$ &$(-25.9\pm 3.2)\times 10^{-3}$\\
GR $\dot{P}(10^{-12}~\rm{s~s^{-1}})$ &$-2.40263\pm 0.00005$&$-27.7^{+1.5}_{-1.9}\times 10^{-3}$\\
 \hline
\end{tabular}
}
\caption{\label{tableI}Summary of the measured orbital parameters and the observed orbital period derivatives, together with the general relativistic predictions, for PSR B1913+16 \cite{Weisberg:2016jye} and PSR J1738+0333 \cite{Freire:2012mg}. The quoted uncertainties correspond to the last significant digits shown in parentheses.}
\end{table}

Hereafter, in ghostly or ghost-free fourth-order gravity, whether from the new force, radiation, or their combined effects, we work under the simplifying assumption $m_{0+}=m_{2+}$. While these parameters could in principle differ, varying them does not significantly affect the observables or the resulting upper limits on the couplings for ghost-free gravity. The essential point is that the radiation from a quasi-stable binary orbit occurs only if the mode masses satisfy $m_{0+}, m_{2+} \lesssim  \Omega$, while the new-force contribution is active only when $m_{0+}, m_{2+}\lesssim 1/a$.

\subsection{Constraints on the standard fourth-order gravity theory from orbital period loss of binary systems}

In the following, we employ the Hulse-Taylor binary and PSR J1738+0333 to constrain the masses of the massive spin-$2$ ghost and spin-$0$ scalar modes using orbital period decay measurements. We present separately the contributions of each mode to the orbital period loss, as well as the effects arising from the modified force, radiation, and their combined influence on the decay rate.

\subsubsection{Orbital period loss of HT binary}
\begin{figure}[htbp]
\centering
\begin{subfigure}[b]{0.4\textwidth}
  \includegraphics[width=\linewidth]{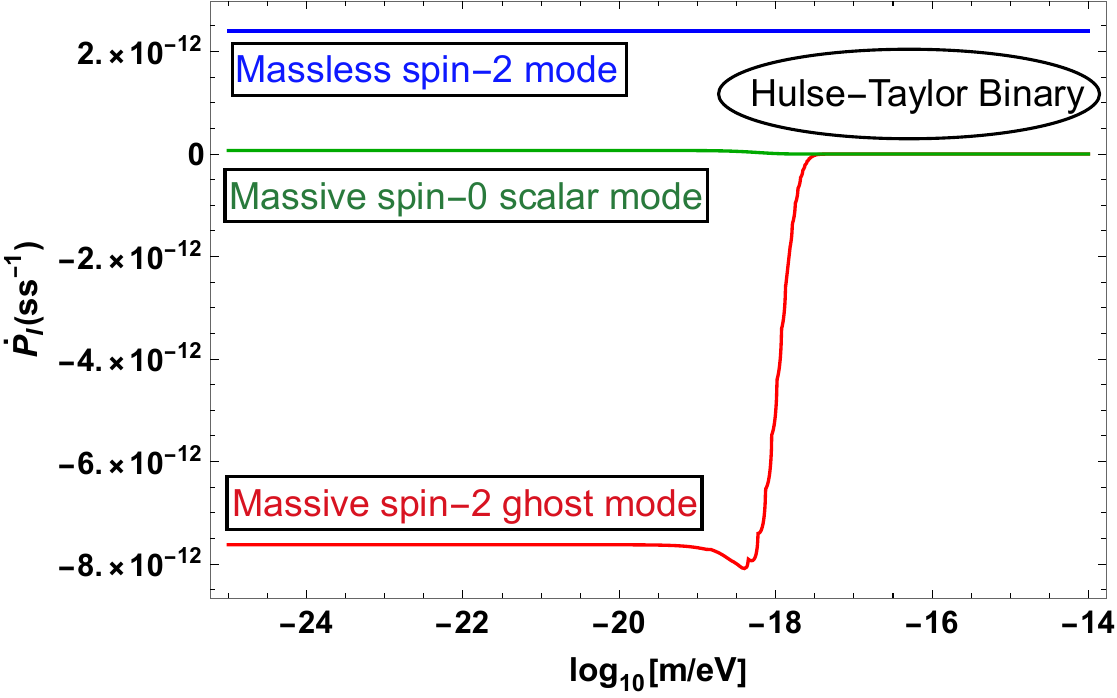}
  \caption{Rates of orbital period loss for individual modes}
  \label{fig:one_a}
\end{subfigure}
\begin{subfigure}[b]{0.4\textwidth}
  \includegraphics[width=\linewidth]{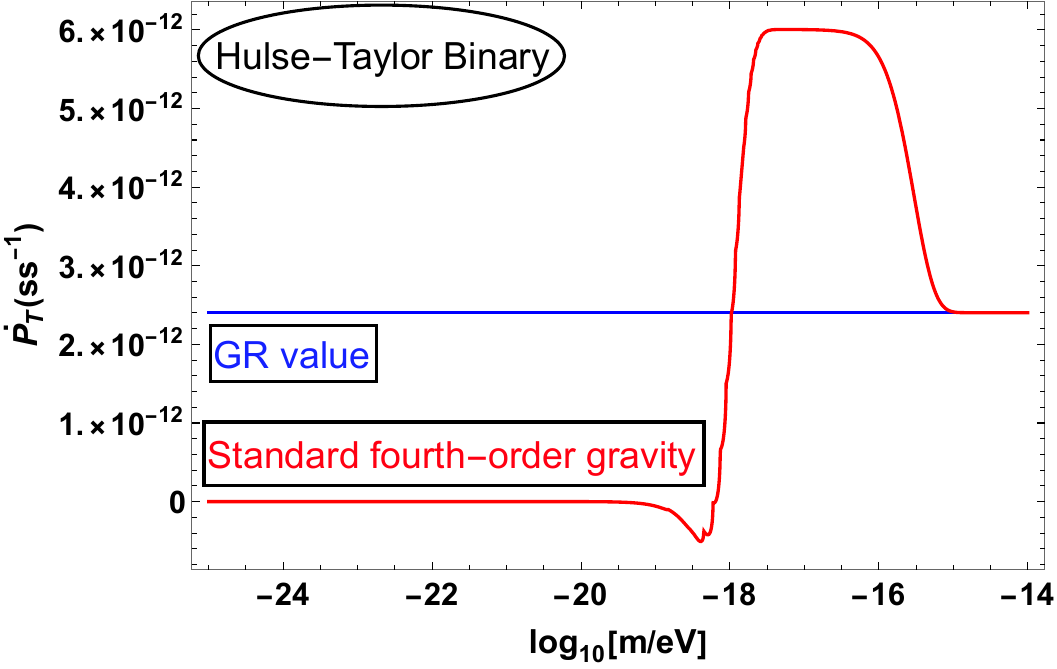}
  \caption{Total rate of orbital period loss compared to the GR value}
  \label{fig:one_b}
\end{subfigure}
\begin{subfigure}[b]{0.4\textwidth}
  \includegraphics[width=\linewidth]{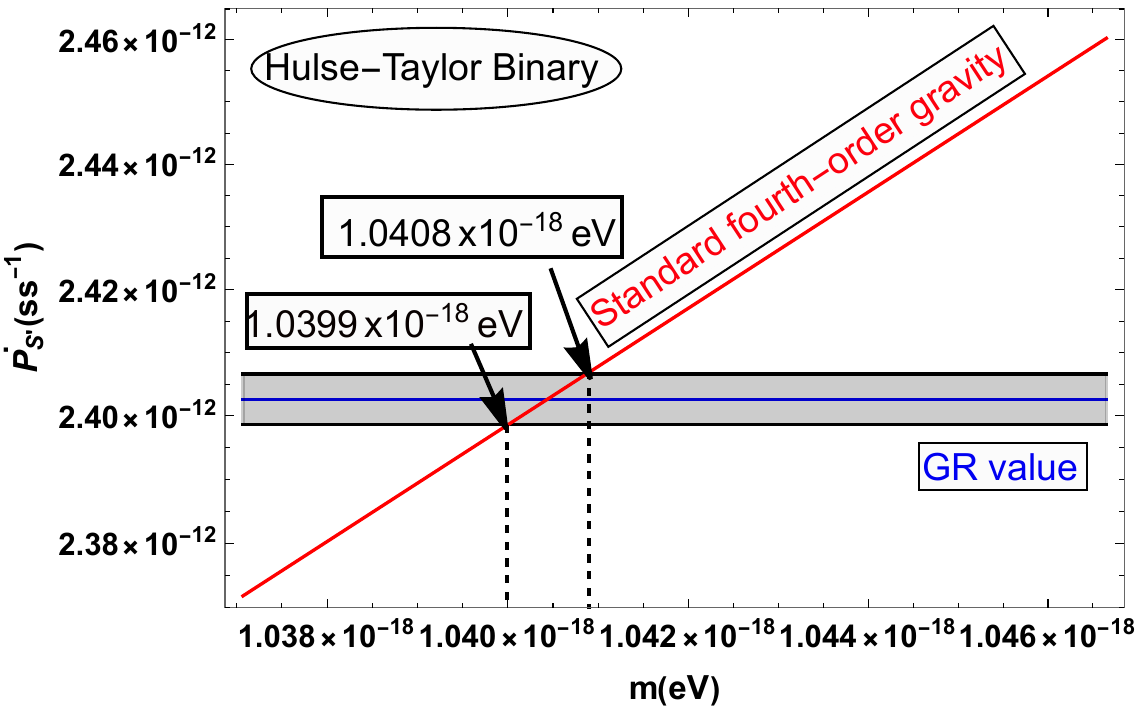}
  \caption{Total rate of orbital period loss in the limit $m\gtrsim \Omega$}
  \label{fig:one_c}
\end{subfigure}
\begin{subfigure}[b]{0.4\textwidth}
  \includegraphics[width=\linewidth]{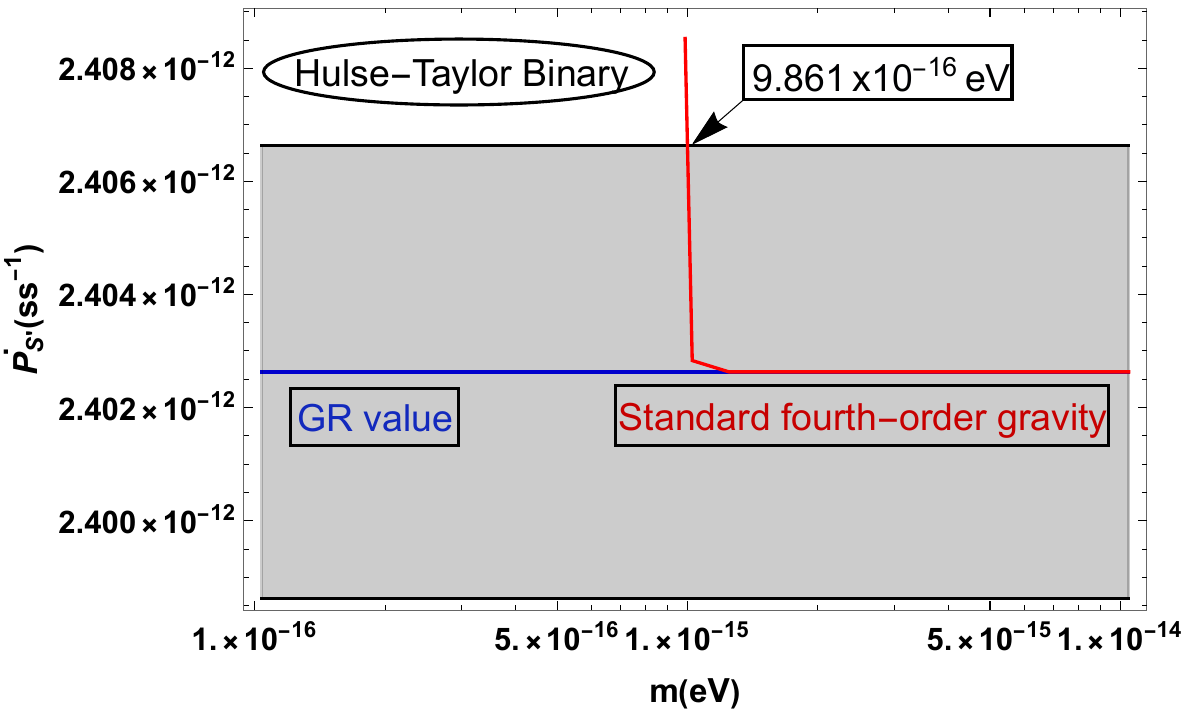}
  \caption{Total rate of orbital period loss in the limit $m\gtrsim 1/a$}
  \label{fig:one_d}
\end{subfigure}

\caption{Contributions of different massless (spin-$2$) and massive (spin-$2$ ghost and spin-$0$) modes, their total contribution and the behaviour in the limit of larger mode masses in standard fourth-order gravity for the rate of orbital period loss of Hulse-Taylor compact binary system: (a) Rates of orbital period loss for the individual modes, (b) Total rate of orbital period loss compared to the GR value, (c) Total rate of orbital period loss in the limit $m\gtrsim \Omega$, and (d) Total rate of orbital period loss in the limit $m\gtrsim 1/a$. See texts for details.}
\label{fig:one}
\end{figure}

In FIG. \ref{fig:one}, we present the separate contributions of individual modes to the orbital period decay rate, together with the total contribution, for the Hulse-Taylor binary system within the framework of standard fourth-order gravity. We also display the rate of orbital period decay in the limit of mode masses greater than the orbital angular frequency and the inverse of the binary separation. The results are obtained using Eqs. \ref{eq:22}, \ref{eq:30}, \ref{eq:39}, and \ref{n5a}. The contributions arise from the massless spin-$2$ mode, the massive spin-$2$ ghost, and the massive spin-$0$ scalar mode.

FIG. \ref{fig:one_a} shows the individual contributions of each mode to the orbital period decay rate $\dot{P}_{I}$, arising from both radiation and the modified force. The effect of a given mode can be isolated by taking the other modes to be infinitely heavy. Due to the additional negative sign in the expression for the orbital period loss associated with the massive spin-2 ghost, its contribution (red curve) appears on the opposite side of the origin compared to the massless spin-2 (blue curve) and the massive spin-0 scalar (green curve). For both massive modes, when their masses exceed the orbital frequency, radiation becomes kinematically forbidden, producing kinks at the corresponding points, followed by a rapid fall-off. The contribution of radiation dominates over that of the modified force in determining the orbital period decay rate from the individual modes. Since the contributions of the massive spin-$2$ ghost and spin-$0$ scalar modes to the orbital period decay approach zero for $m\gtrsim \Omega$, the modifications to the force law become negligible in this regime. Furthermore, when both massive modes are made infinitely heavy, the massless mode (blue curve) alone recovers the standard GR prediction for the orbital period loss, as expected.

FIG. \ref{fig:one_b} illustrates the total orbital period loss rate (red curve) $\dot{P}_T$, which is compared with the GR prediction (blue curve). In the standard fourth-order gravity scenario, GW radiation is completely suppressed, resulting in the red curve approaching zero for small values of the masses, $m_{2+}, m_{0+} \lesssim \Omega \sim 1.48 \times 10^{-19}~\mathrm{eV}$ for HT binary. This cancellation arises because the combined effect of the massive spin-$2$ ghost and spin-$0$ scalar modes in measuring the GW flux negates the contribution from the massless spin-$2$ mode.

Note, this cancellation does not mean that there is no emission of GWs. Indeed, due to this cancellation, any source of GWs can emit a pair of positive and negative energy waves, and such emission can be repeated without costing the energy of the source. Hence, at the end, the universe would be filled with ordinary and ghostly GWs. At the quantum level, the pair emission can be triggered even by quantum fluctuations at any place and at any time at arbitrarily shorter scales all the way down to the cutoff length of the theory, leading to fatal instability of the vacuum.

Notably, when neither the massive spin-$2$ nor spin-$0$ modes are radiated, the massless spin-$2$ mode alone does not reproduce the GR result for $m\lesssim 1/a$. This is due to the deviation of the gravitational potential in fourth-order gravity from the standard Newtonian form. As a consequence, the red curve in FIG. \ref{fig:one_b} crosses above the blue curve, indicating a larger orbital period loss than predicted by GR for larger values of the mode masses ($m\gtrsim \Omega$). The factor of $2.5$ enhancement in the orbital period loss relative to the standard GR result arises from the first bracketed term of Eq. \ref{n5a}, which originates from the modified force law. Near the threshold $m_{0+}, m_{2+} \sim \Omega$, a dip appears in the red curve, which originates from the kinematic (phase-space) suppression of the emitted radiation. An additional kink arises when $m_{0+}, m_{2+} \sim 1/a$, stemming from the modified form of the force law. In the regime $m_{0+}, m_{2+} \gtrsim 1/a$, and consequently $m_{0+}, m_{2+} \gtrsim \Omega$, the modifications to both radiation and the force law become suppressed, and standard fourth-order gravity effectively reduces to Einstein's GR. Thus, the red curve coincides with the blue curve, corresponds to the GR result.

In FIG. \ref{fig:one_c}, the gray band represents the allowed $1\sigma$ uncertainty of the orbital period loss measurement from observations. For mode masses $m\gtrsim \Omega$, the red curve, which is the total orbital period loss ($\dot{P_{S^\prime}}$) crosses this band, implying that the allowed mass range for the massive modes is constrained to
$1.0399 \times 10^{-18}~\mathrm{eV} \lesssim m_{0+}, m_{2+} \lesssim 1.0408 \times 10^{-18}~\mathrm{eV}$.

In FIG. \ref{fig:one_d}, the gray band carries the same interpretation as in FIG. \ref{fig:one_c}. For mode masses $m\gtrsim 1/a$, the total orbital period loss (red curve, $\dot{P_{S^\prime}}$) intersects this band, indicating that the viable parameter space for the massive modes is restricted to $m_{0+}, m_{2+}\gtrsim 9.861\times 10^{-16}~\mathrm{eV}$.  

\subsubsection{Orbital period loss of PSR J1738+0333 binary}
\begin{figure}[htbp]
\centering
\begin{subfigure}[b]{0.4\textwidth}
  \includegraphics[width=\linewidth]{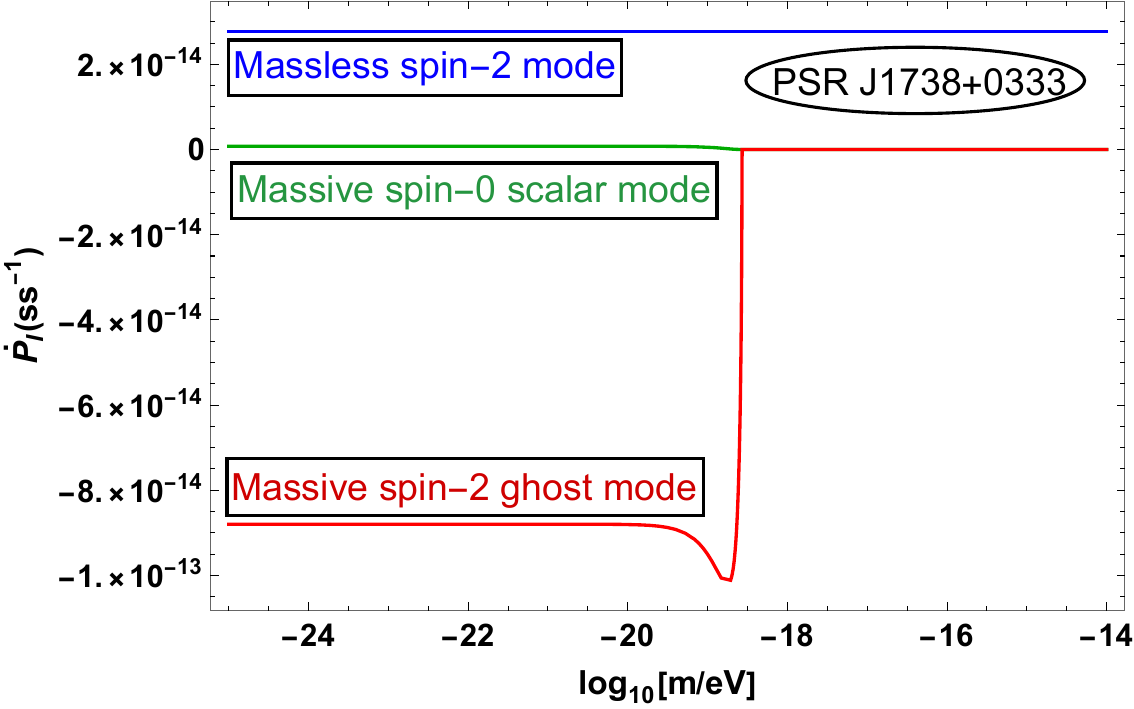}
  \caption{Rates of orbital period loss for individual modes}
  \label{fig:two_a}
\end{subfigure}
\begin{subfigure}[b]{0.4\textwidth}
  \includegraphics[width=\linewidth]{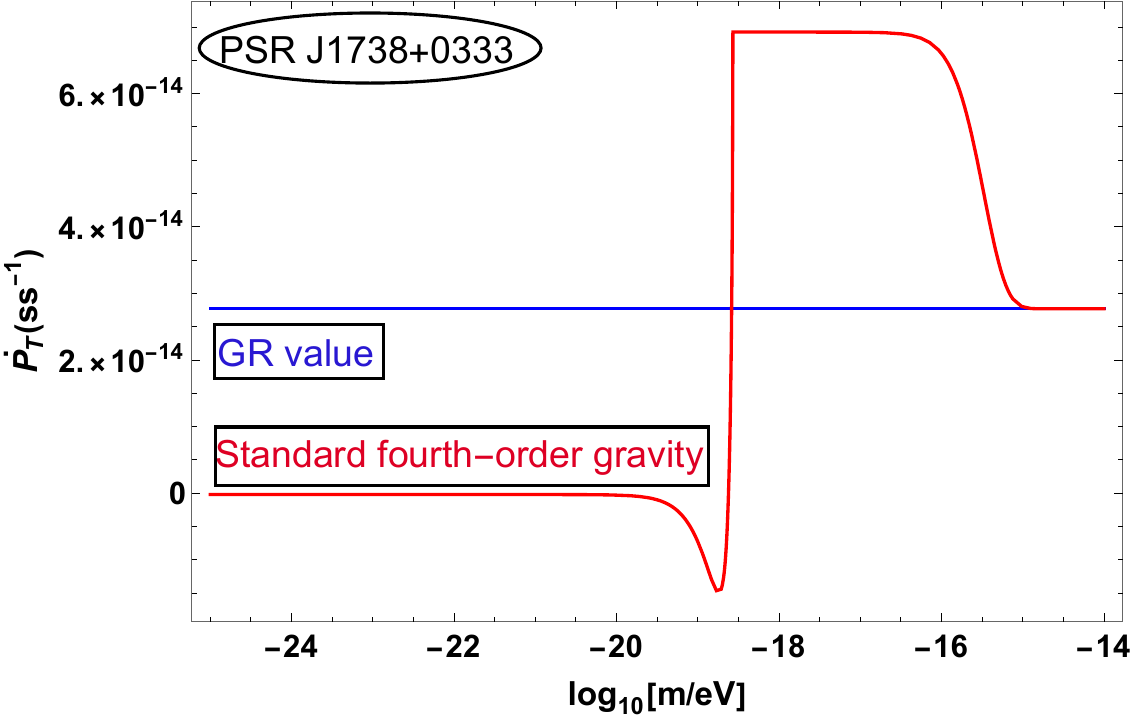}
  \caption{Total rate of orbital period loss compared to the GR value}
  \label{fig:two_b}
\end{subfigure}
\begin{subfigure}[b]{0.4\textwidth}
  \includegraphics[width=\linewidth]{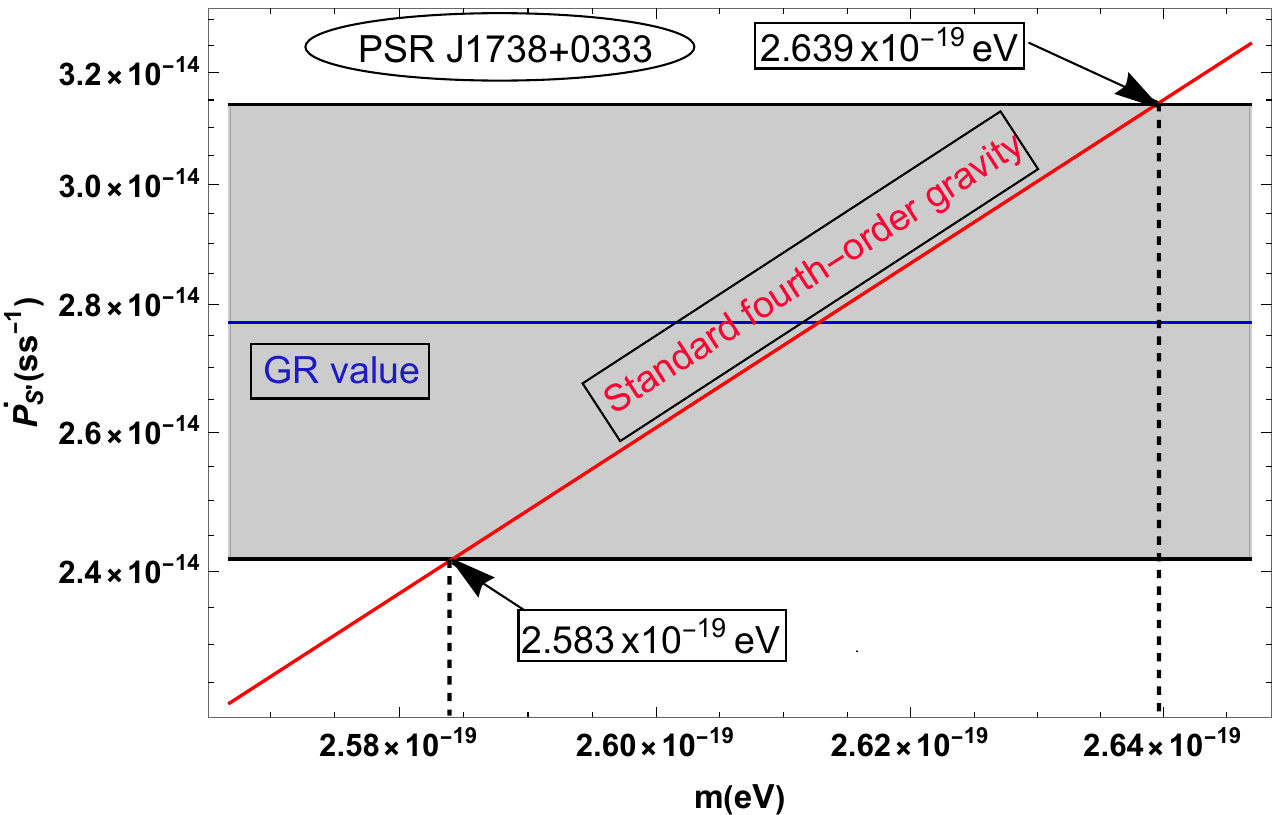}
  \caption{Total rate of orbital period loss in the limit $m\gtrsim \Omega$}
  \label{fig:two_c}
\end{subfigure}
\begin{subfigure}[b]{0.4\textwidth}
  \includegraphics[width=\linewidth]{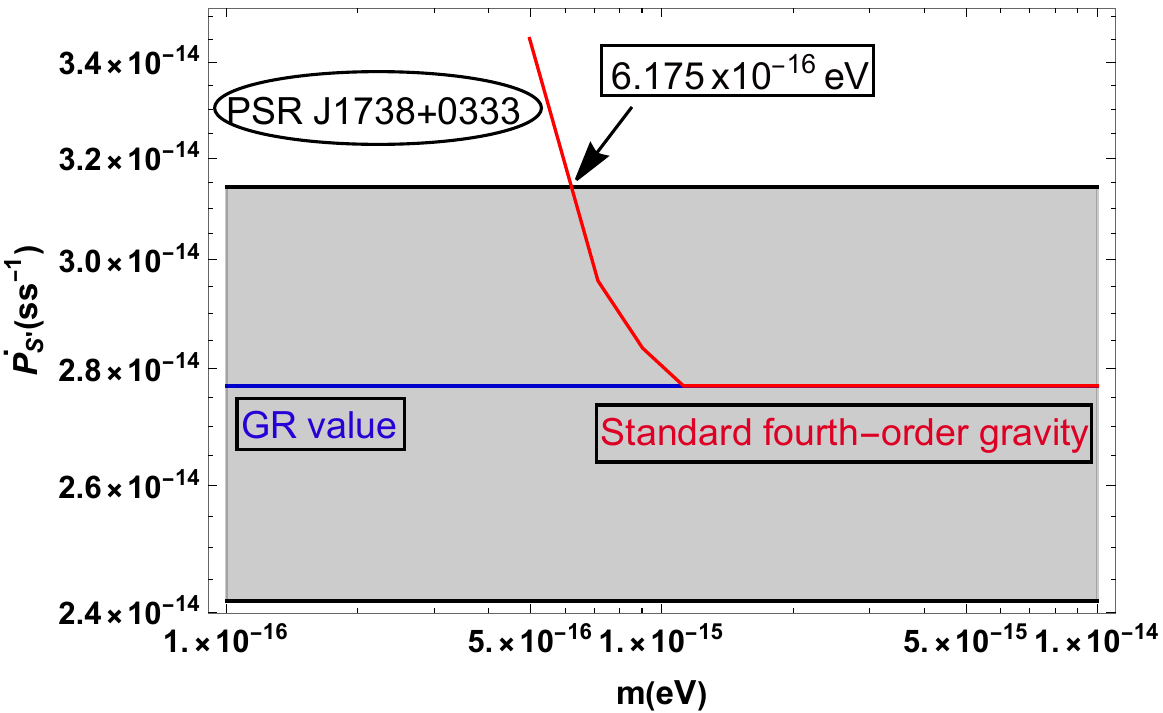}
  \caption{Total rate of orbital period loss in the limit $m\gtrsim 1/a$}
  \label{fig:two_d}
\end{subfigure}

\caption{Contributions of different massless (spin-$2$) and massive (spin-$2$ ghost and spin-$0$) modes and their total contribution in standard fourth-order gravity for the rate of orbital period loss of PSR J1738+0333 compact binary system: (a) Rates of orbital period loss for the individual modes, (b) Total rate of orbital period loss compared to the GR value, (c) Total rate of orbital period loss in the limit $m\gtrsim \Omega$, and (d) Total rate of orbital period loss in the limit $m\gtrsim 1/a$. See texts for details.}
\label{fig:two}
\end{figure}
In FIG. \ref{fig:two}, we present the same analysis as in FIG. \ref{fig:one}, but applied to PSR J1738+0333. Particularly, in FIG. \ref{fig:two_a}, we show the separate contributions of the three modes in the orbital period decay rate; in FIG. \ref{fig:two_b}, the total orbital period loss compared with the GR prediction; in FIG. \ref{fig:two_c}, the total orbital period loss in the limit $m\gtrsim \Omega$; and in FIG. \ref{fig:two_d}, the total orbital period loss rate in the limit $m\gtrsim 1/a$. The behaviour of the individual modes are exactly same as for HT binary except the numericals are different due to different orbital parameters (FIG. \ref{fig:two_a}). Within the standard fourth-order gravity framework, GW emission is completely suppressed, which causes the red curve to approach zero for small masses, $m_{2+}, m_{0+} \lesssim \Omega \sim 1.35 \times 10^{-19}~~\mathrm{eV}$ for PSR J1738+0333 (FIG. \ref{fig:two_b}). For mode masses $m\gtrsim \Omega$, the red curve intersects the allowed $1\sigma$ uncertainty band for the measurement of orbital period loss of PSR J1738+03333, restricting the allowed range of the massive modes to $2.583 \times 10^{-19}~\mathrm{eV} \lesssim m_{0+}, m_{2+} \lesssim 2.639 \times 10^{-19}~\mathrm{eV}$ (FIG. \ref{fig:two_c}). In the limit $m_{0+}, m_{2+}\gtrsim 1/a$, the red curve intersects this band, indicating that the viable parameter space for the massive modes is restricted to $m_{0+}, m_{2+}\gtrsim 6.175\times 10^{-16}~\mathrm{eV}$ (FIG. \ref{fig:two_d}). Since the mass ranges obtained from the Hulse–Taylor binary and PSR J1738+0333 do not overlap in the regime $m\lesssim \Omega$, no universal upper bound on the massive modes can be established in that limit. In the regime $m\gtrsim 1/a$, the tightest lower limit on the massive modes from both binary systems is found to be $m\gtrsim 6.175\times 10^{-16}~\mathrm{eV}$.

\subsubsection{Orbital period loss in different limiting cases for HT and PSR J1738+0333}
\begin{figure}[htbp]
\centering
\begin{subfigure}[b]{0.4\textwidth}
  \includegraphics[width=\linewidth]{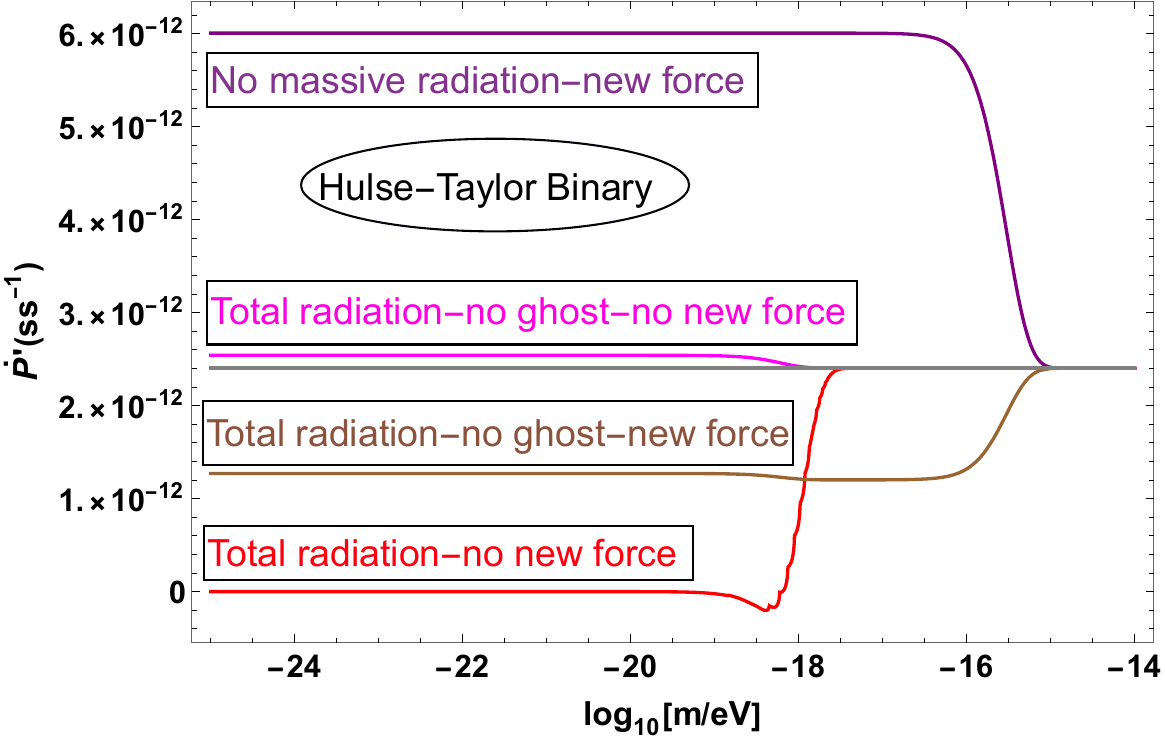}
  \caption{Orbital period loss in different limiting cases for Hulse-Taylor binary}
  \label{fig:three_a}
\end{subfigure}
\begin{subfigure}[b]{0.4\textwidth}
  \includegraphics[width=\linewidth]{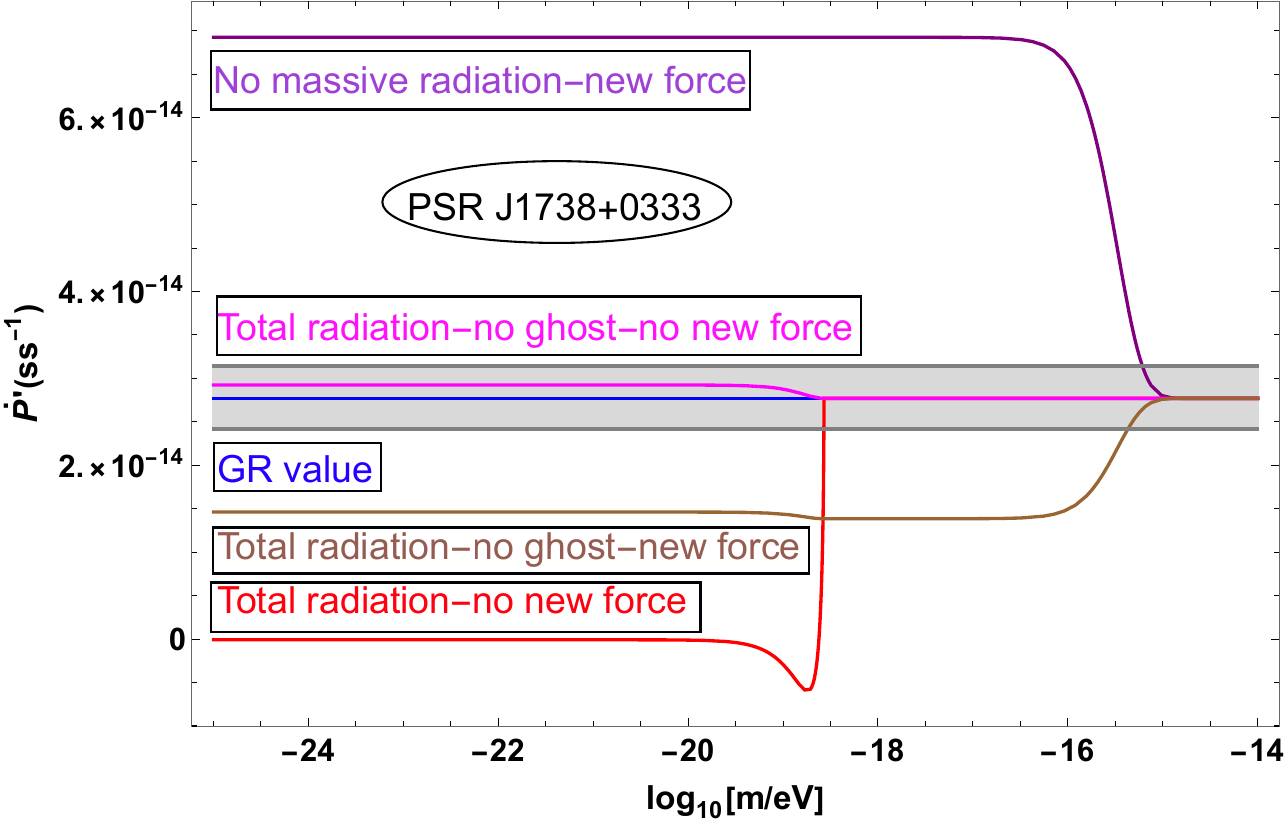}
  \caption{Orbital period loss in different limiting cases for PSR J1738+0333}
  \label{fig:three_b}
\end{subfigure}

\caption{Orbital period loss in different limiting cases for (a) Hulse-Taylor binary and (b) PSR J1738+0333 in standard fourth-order gravity. See texts for details.}
\label{fig:three}
\end{figure}
In FIG. \ref{fig:three}, we present the orbital period loss, $\dot{P^\prime}$ for different limiting cases of the Hulse–Taylor binary (FIG. \ref{fig:three_a}) and PSR J1738+0333 (FIG. \ref{fig:three_b}). There are two characteristic length (or energy) scales-the orbital separation between the two stars and the orbital frequency of the binary. We use Eqs. \ref{eq:22}, \ref{eq:30}, \ref{eq:39} and \ref{n5a} in plotting FIG. \ref{fig:three}.

The new massive scalar and spin-2 modes mediate long-range forces when their masses satisfy $m_{2+}, m_{0+}\lesssim 1.01\times 10^{-16}~\mathrm{eV}$ (correspond to the inverse of the semi-major axis $a$ of the orbit) for the Hulse-Taylor binary system and $m_{2+}, m_{0+}\lesssim 1.14\times 10^{-16}~\mathrm{eV}$ for PSR J1738+0333. Moreover, these modes can be radiated from the binary if their masses are smaller than the orbital frequency ($\Omega$), i.e., $m_{2+}, m_{0+}\lesssim 1.48\times 10^{-19}~\mathrm{eV}$ for the Hulse-Taylor system and $m_{2+}, m_{0+}\lesssim 1.35\times 10^{-19}~\mathrm{eV}$ for PSR J1738+0333.

Thus, when $m_{2+}, m_{0+}>\Omega$ but $m_{2+}, m_{0+}<1/a$, the massive modes do not radiate but still generate an additional long-range force that modifies the orbital period decay. In contrast, when $m_{2+}, m_{0+}<1/a$ and simultaneously $m_{2+}, m_{0+}<\Omega$, both new-force effects and radiation from the massive modes contribute to the orbital period loss. 

In FIGs. \ref{fig:three_a} and \ref{fig:three_b}, the purple curves show the orbital period decay rate when massive-mode radiation is absent i.e., $m=\{m_{2+}, m_{0+}\}> \Omega$, while the new-force correction induces a distinct bending of the curve for $m < 1/a$ $(m=\{m_{2+}, m_{0+}\})$. Due to this modification, the purple curve does not approach the GR result as $m \to 0$, and the effect enhances the orbital period decay by a factor of about 2.5 relative to the GR prediction.

If the ghost mode is consistently integrated out by making it infinitely heavy, only the massive scalar together with the usual massless spin-2 graviton remain dynamical. The brown curves in FIGs. \ref{fig:three_a} and \ref{fig:three_b} correspond to this scenario, where the orbital period loss arises solely from the combined effects of the modified force and radiation due to the massless spin-2 and the massive scalar modes and the ghost is infinitely heavy to be integrated out. Therefore, there is a bending of the curve at $1/a$ and $\Omega$, coming from the contributions of modified force law and radiation respectively.

The red and magenta curves in FIGs. \ref{fig:three_a} and \ref{fig:three_b} represent the contribution to orbital period loss arising solely from radiation, in the absence of any new force. The magenta curve additionally corresponds to the case where the ghost contribution in radiation is absent (integrated out). Similar to the case with a new force present, when the new force is switched off, the radiation alone does not yield any measurable contribution to the orbital period loss. The new force is inactive when the massive modes do not act as mediators, and radiation can only occur if the modes go on shell, which requires $m \lesssim \Omega$ ($m=\{m_{2+}, m_{0+}\}$). Due to the relatively large uncertainties in the orbital period loss measurements, the massive scalar mode can still be constrained from PSR J1738+0333 in the absence of the ghost (magenta curve for PSR J1738+0333). In contrast, the more precise measurements from the Hulse-Taylor binary do not yield any limit on the mass of the scalar mode and hence no universal bound on the scalar mode mass is obtained for $m\lesssim \Omega$.

\subsection{Constraints on the ghost-free quadratic gravity theory from orbital period loss of binary systems}
\begin{figure}[htbp]
\centering
\begin{subfigure}[b]{0.41\textwidth}
  \includegraphics[width=\linewidth]{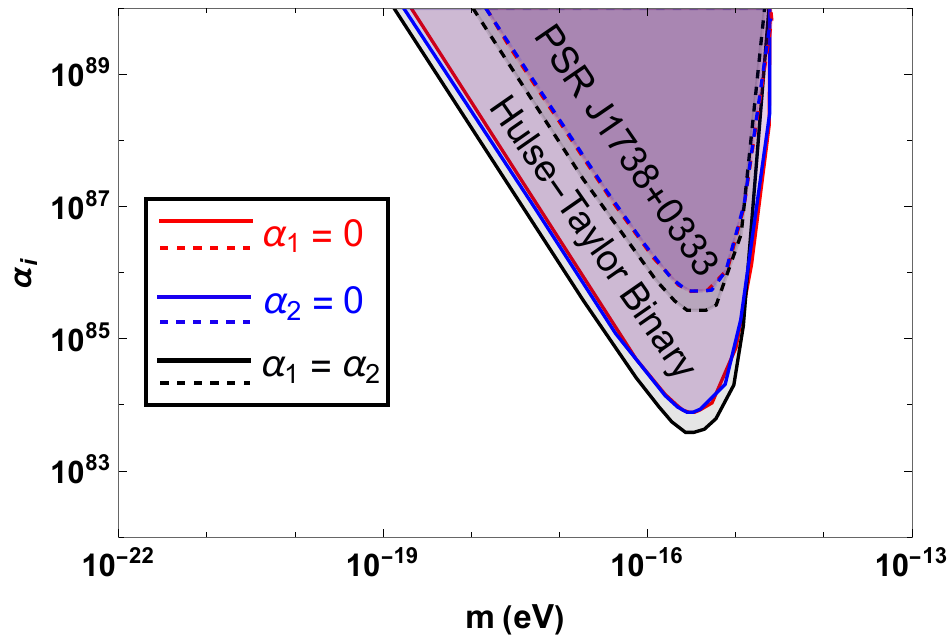}
  \caption{Constraints on coupling from new force+radiation effects}
  \label{fig:sm1}
\end{subfigure}
\begin{subfigure}[b]{0.4\textwidth}
  \includegraphics[width=\linewidth]{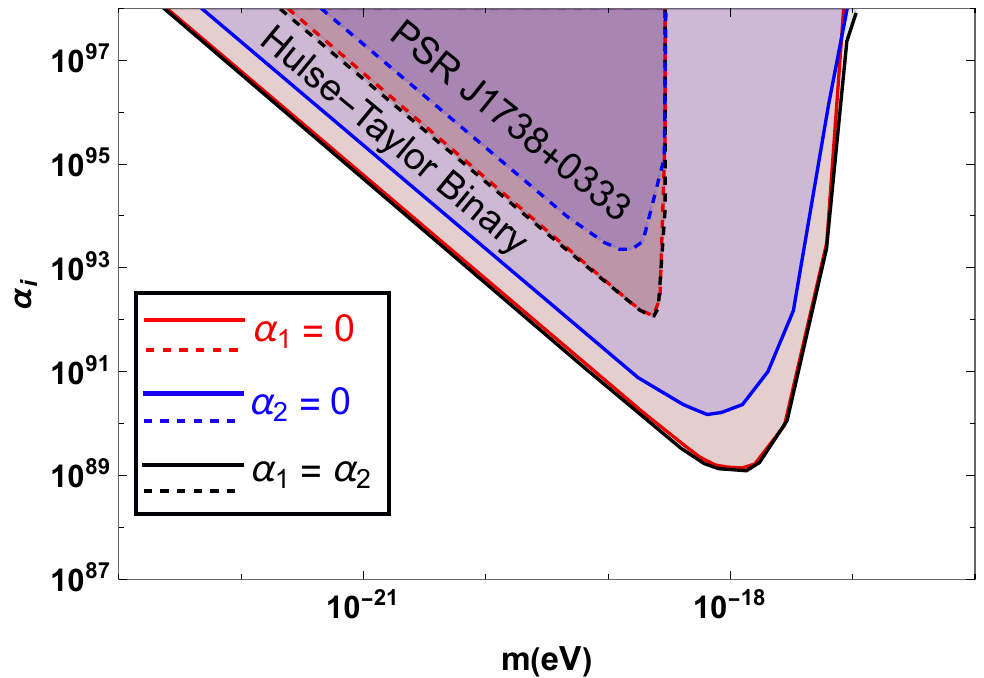}
  \caption{Constraints on coupling from radiation effect only}
  \label{fig:sm2}
\end{subfigure}
\caption{Constraints on the couplings $\alpha_i$ for ghost-free fourth-order gravity from orbital period decay measurements of the Hulse-Taylor binary and PSR J1738+0333, considering (a) the combined effects of the modified force and additional radiation, and (b) radiation effects only. See texts for details.}
\label{fig:sm}
\end{figure}

In FIG. \ref{fig:sm}, we present the bounds on the couplings $\alpha_i~(\alpha_1, \alpha_2)$ of fourth-order ghost-free gravity derived from orbital period loss measurements of the Hulse-Taylor binary and PSR J1738+0333. We use Eqs. \ref{eq:22}, \ref{eq:28}, \ref{eq:37}, and \ref{n5} in obtaining FIG. \ref{fig:sm}. Constraints on the couplings $\alpha_i$ are derived for two scenarios: when both the modified force and radiation contribute to the orbital period decay (FIG. \ref{fig:sm1}), and when the decay is driven solely by radiation (FIG. \ref{fig:sm2}). Among these systems, the Hulse-Taylor binary provides the tighter constraints. The new-force effects dominate near the mass scale $m \sim 1/a$, whereas for $m \ll 1/a$ the modification becomes indistinguishable from Newtonian gravity, leading to weaker bounds. Similarly, for $m \gg 1/a$, the long-range force approximation ceases to hold. The radiation effect is most significant near $m \sim \Omega$, but becomes suppressed for $m \gg \Omega$. In the opposite limit, $m \ll \Omega$, the massive modes make no contribution since the masses of the modes enter in the numerator of the orbital period decay expression. In deriving the bounds on the coupling, we assume $m=m_{2+}=m_{0+}$. The red curves ($\alpha_1=0$) represent the scenario where both massive modes contribute, but with reduced strength compared to the $\alpha_1=\alpha_2$ case. The blue curves ($\alpha_2=0$) show the situation where the massive spin-2 mode contribution vanishes. The black curves correspond to the case $\alpha_1=\alpha_2$. Solid lines indicate results from the Hulse-Taylor binary, while dashed lines denote those from PSR J1738+0333.

The strongest limits arise when both couplings $\alpha_1$ and $\alpha_2$ contribute equally. If one of them vanishes, the bounds become comparatively weaker, as expected, as the effects are additive. In particular, setting $\alpha_2 = 0$ removes the contribution of the massive spin-2 mode, while setting $\alpha_1 = 0$ leaves contributions from both the massive spin-2 and spin-0 modes, though at a reduced strength compared to the $\alpha_1=\alpha_2$ case. Consequently, in FIG. \ref{fig:sm1}, the most stringent bound is obtained for $\alpha_1 \simeq \alpha_2 \lesssim 4.13\times 10^{83}$ at a characteristic mass scale $m_{0+} \sim m_{2+} \sim 3.1\times 10^{-16}~\mathrm{eV}$, from Hulse-Taylor binary system, when both modified force and radiation contribute to the orbital period loss.

In FIG. \ref{fig:sm2}, we show the constraints on the couplings $\alpha_{1,2}$ under the assumption that the orbital period loss arises solely from radiation. This corresponds to the scenario where the massive modes do not mediate a new-force but can be emitted as on-shell radiation. The HT binary provides the strongest constraints, with the most stringent bound obtained for $\alpha_1 \simeq \alpha_2 \lesssim 1.4\times 10^{89}$ at a characteristic mass scale $m_{0+}\sim m_{2+}\sim 1.2\times 10^{-18}~~\mathrm{eV}$.

Therefore, in standard fourth-order gravity no universal upper bounds on the mass parameters can be derived from the orbital period loss of quasi-stable binaries, which are sensitive to scales below $10^{-16}~\mathrm{eV}$ and $10^{-18}~\mathrm{eV}$. Although, the lower bounds on the mass of the modes are obtained as $m\gtrsim 6.175\times 10^{-16}~\mathrm{eV}$. In contrast, ghost-free fourth-order gravity does yield upper bounds on the couplings at the mass scales of $10^{-16}~\mathrm{eV}$ and $10^{-18}~\mathrm{eV}$, which are $\alpha_1 \simeq \alpha_2 \lesssim 4.13\times 10^{83}$ at $m_{0+} \sim m_{2+} \sim 3.1\times 10^{-16}~\mathrm{eV}$ and $\alpha_1 \simeq \alpha_2 \lesssim 1.4\times 10^{89}$ at $m_{0+}\sim m_{2+}\sim 1.2\times 10^{-18}~~\mathrm{eV}$. 

In the next section, we extend the analysis to shorter length scales by deriving constraints on the mass parameters and couplings from the coalescing binary GW170817, which makes the binary stars to come close to each other and increases the orbital frequency.

\section{Constraints from GW170817}\label{sec8}
In this section, we derive constraints on the mass parameters and couplings from the GW170817 event within both standard and ghost-free fourth-order gravity. Unlike the previous case, GW170817 allows us to probe these theories at shorter length scales as small as of order $\sim 20~\mathrm{km}$.

\subsection{Constraints on the standard fourth-order gravity from GW170817}
\begin{figure}
    \centering
    \includegraphics[width=0.5\linewidth]{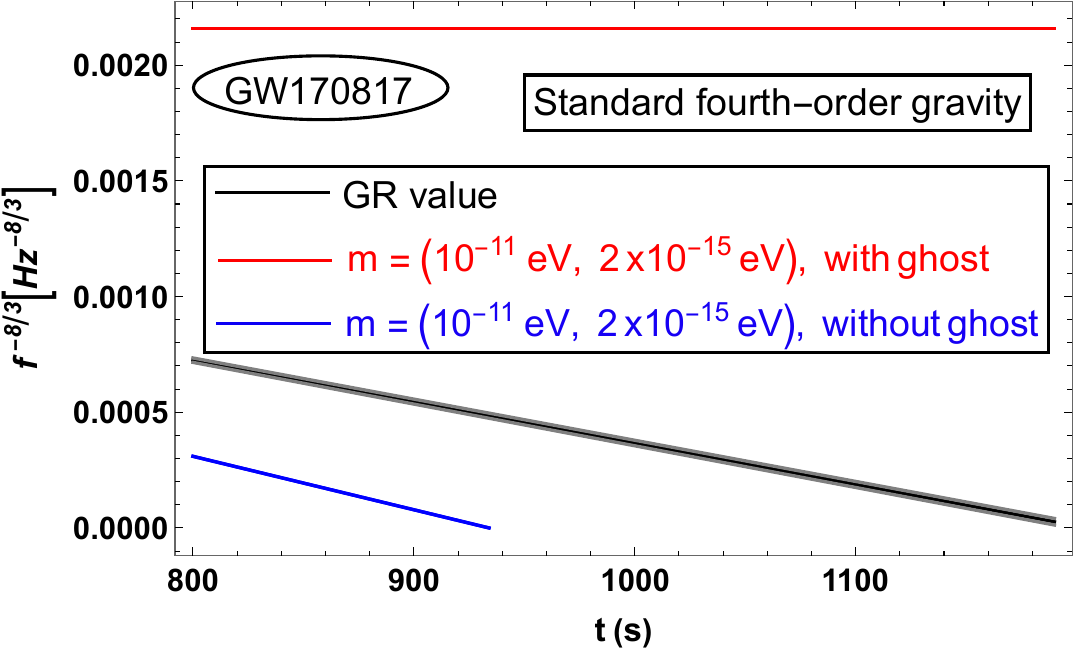}
    \caption{Constraints on the mass of the modes from GW170817 in standard fourth-order gravity. See texts for details.}
    \label{fig:gwfourth-order}
\end{figure}

In FIG. \ref{fig:gwfourth-order}, we show the variation of the GW frequency for GW170817 as a function of the binary NS coalescence time. We use Eq. \ref{rad10} with the substitutions $\alpha=-4/3$, $\beta=1/3$, $2m^2_{2+}\alpha_2/M^2_{pl}=-1$, $4m^2_{0+}(3\alpha_1+\alpha_2)/M^2_{pl}=1$, and $f=\Omega/\pi$ for the standard fourth-order gravity theory in plotting FIG. \ref{fig:gwfourth-order}. We plot the GW frequency raised to the power $-8/3$, that is $f^{-8/3}$, against the coalescence time, which yields a linear relation and facilitates direct comparison between Einstein's GR and the modified gravity theories. The black line represents the GR prediction, while the grey-shaded region indicates the uncertainty band arising from the measured chirp mass of the event, which we have considered to be $0.4\%$, as a conservative limit. We vary the mass of the additional modes in the range $2\times 10^{-15}~~\mathrm{eV}\leq m \leq 10^{-11}~\mathrm{eV}$, with $m=\{m_{2+}, m_{0+}\}$. 

The variation of the GW frequency with the coalescence time is shown by the red line in FIG. \ref{fig:gwfourth-order}, where all the massive and massless modes contribute. Within the mass range $2\times 10^{-15}~~\mathrm{eV}\leq m \leq 10^{-11}~\mathrm{eV}$, the predicted frequency evolution in fourth-order gravity lies outside the observational window of GW170817. Consequently, the event does not allow us to place bounds on the masses of these modes in this framework. At leading order in the GW energy loss, the contributions from the massive spin-$2$ ghost mode and the massive spin-$0$ scalar mode cancel those of the massless spin-$2$ mode. As a result, the GW frequency remains constant in time, and its value is lower than in GR (since $f^{-8/3}$ is correspondingly larger), reflecting the cancellation of the massless mode contribution. Also, the blue curve in FIG. \ref{fig:gwfourth-order} shows the frequency evolution in the absence of the ghost. In this case, $\dot{\Omega}$ is nonzero, with both the massive scalar and the massless spin-2 mode contributing. Consequently, $f^{-8/3}$ decreases with time, as in GR, but the overall frequency is higher than in GR due to the additional scalar contribution. Consequently, the frequency-chirp measurement of GW170817 event constrains the mass of the modes in ghostful theory as $m\gtrsim 10^{-11}~\mathrm{eV}$.

The chosen mass range is motivated by the GW170817 event, since $m\sim 10^{-11}~\mathrm{eV}$ 
$(m=\{m_{2+}, m_{0+}\})$ corresponds to the twice of the inverse length scale of the NS radius. On the contrary, $10~\mathrm{Hz}$ GW frequency is LIGO threshold, which corresponds to the mass scale $m\sim 6.58\times 10^{-15}~\mathrm{eV}$. Direct GW observations can only constrain mode masses below this value, as the GW wavelength must exceed the stellar size to be detectable.

\subsection{Constraints on the ghost-free fourth-order gravity from GW170817}

In the following, we use the GW170817 event to constrain the couplings and masses of the modes in ghost-free quadratic gravity, based on chirp-mass measurements extracted from the evolution of the GW frequency with coalescence time. We analyze the separate and combined contributions of the modified force and radiation to the orbital frequency evolution, and derive bounds from the frequency-chirp measurements.

\subsubsection{Effects of new force}
\begin{figure}[htbp]
\centering
\begin{subfigure}[b]{0.4\textwidth}
  \includegraphics[width=\linewidth]{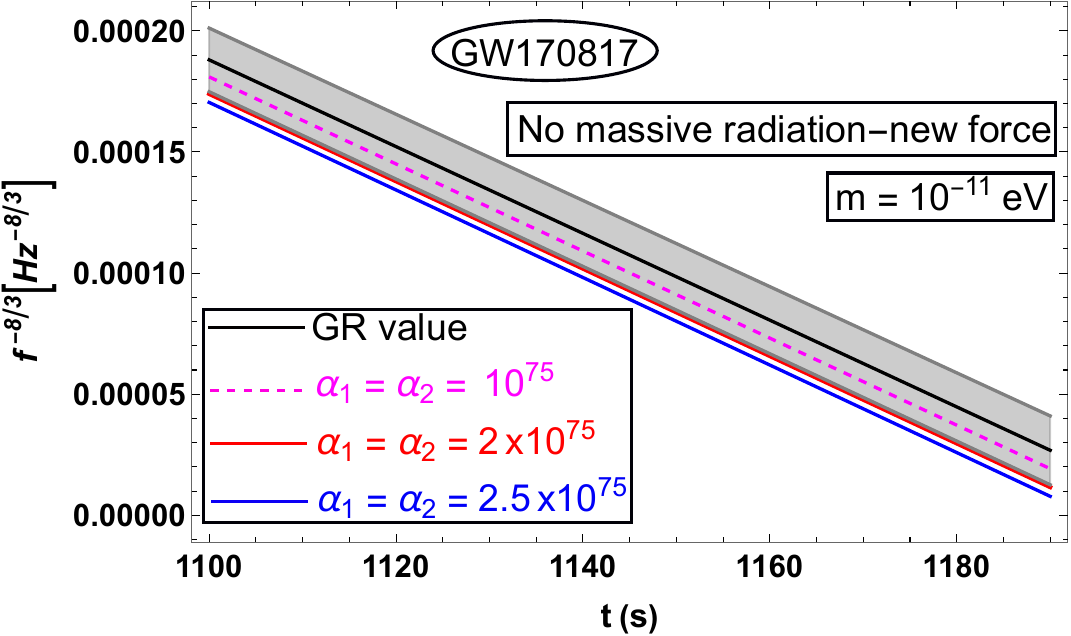}
  \caption{Constraints on couplings from new force only for $m=10^{-11}~\mathrm{eV}$}
  \label{fig:gw1}
\end{subfigure}
\begin{subfigure}[b]{0.4\textwidth}
  \includegraphics[width=\linewidth]{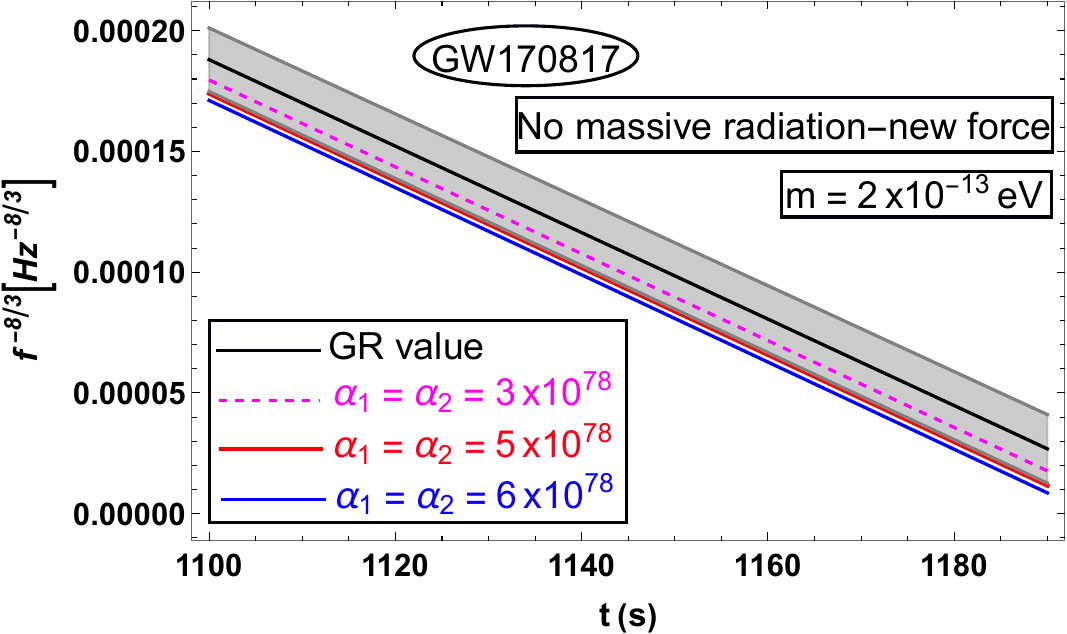}
  \caption{Constraints on couplings from new force only for $m=2\times 10^{-13}~\mathrm{eV}$}
  \label{fig:gw2}
\end{subfigure}
\caption{Constraints on the couplings $\alpha_i$ from the GW170817 event in ghost-free quadratic gravity, including new-force effects only, shown for (a) $m=10^{-11}~~\mathrm{eV}$ and (b) $m=2\times 10^{-13}~~\mathrm{eV}$. See texts for details.
}
\label{fig:gwone}
\end{figure}

In FIG. \ref{fig:gwone}, we present the constraints on the couplings in ghost-free quadratic gravity obtained from the GW170817 chirp-mass measurement in the case where only the new force effect is active, i.e., when the massive modes act as mediators. This situation arises when the mode masses exceed $2\Omega \sim 10^{-14}~\mathrm{eV}$, corresponding to the LIGO threshold frequency. The results are based on Eq. \ref{rad10}, where terms with exponential factors contribute to the orbital frequency evolution, while those involving $(1 - m^2/4\Omega^2)$ are suppressed. 

The frequency evolution is computed as a function of the binary coalescence time for representative masses $m=10^{-11}~\mathrm{eV}$ (FIG. \ref{fig:gw1}) and $m=2\times 10^{-13}~\mathrm{eV}$ (FIG. \ref{fig:gw2}), corresponding to binary separations in the range $20~\mathrm{km} \lesssim r \lesssim 750~\mathrm{km}$, where the GW signal enters the LIGO/Virgo sensitivity band. The evolution is shown for different values of the couplings $\alpha_i~(i=1,2)$. For $\alpha_i \sim 2.5\times 10^{75}\,(6\times 10^{78})$, the predicted GW frequency falls outside the observational gray-shaded region for $m\sim 10^{-11}\,(2\times 10^{-13})~\mathrm{eV}$, as indicated by the blue curves. Since no significant deviations from GR are observed, we derive upper bounds $\alpha_i \lesssim 2\times 10^{75}$ for $m\sim 10^{-11}~\mathrm{eV}$ and $\alpha_i \lesssim 5\times 10^{78}$ for $m\sim 2\times 10^{-13}~\mathrm{eV}$, shown by the red curves. The magenta dashed curves lie comfortably within the observational band, corresponding to $\alpha_i\sim 10^{75}\,(3\times 10^{78})$ for $m\sim 10^{-11}\,(2\times 10^{-13})~\mathrm{eV}$. The bounds strengthen as the mode masses increase.

The inferred coalescence time is shifted by $\mathcal{O}(7.3~\mathrm{s})$ relative to GR for $\alpha_i\sim 2\times 10^{75}$ at $m=10^{-11}~\mathrm{eV}$, and for $\alpha_i\sim5\times 10^{78}$ at $m=2\times 10^{-13}~\mathrm{eV}$. Higher-order corrections to GW emission have not been included in this analysis; such terms could break degeneracies between new force effects and GR in regimes where the Yukawa potential of the new interaction is unsuppressed within the observational window. A more robust analysis would require generating full waveform templates including these corrections and comparing them to detector noise curves.

\subsubsection{Effects of radiation}

\begin{figure}[htbp]
\centering
\begin{subfigure}[b]{0.4\textwidth}
  \includegraphics[width=\linewidth]{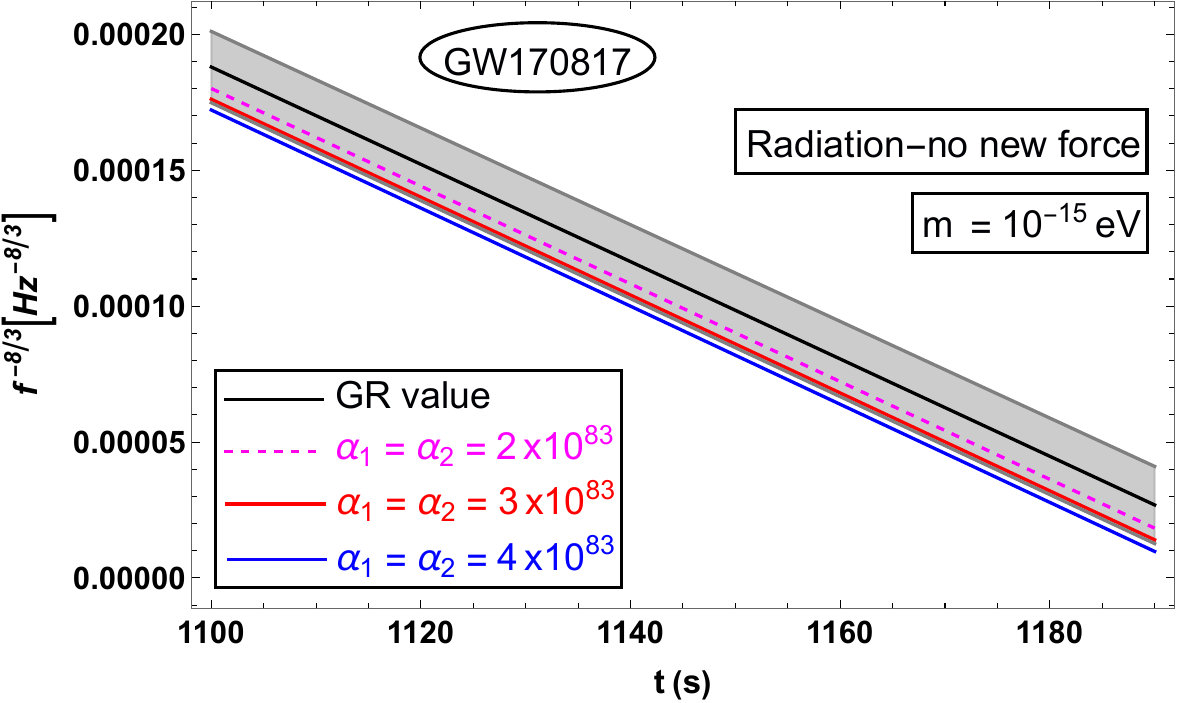}
  \caption{Constraints on couplings from radiation only for $m= 10^{-15}~\mathrm{eV}$}
  \label{fig:gwtwoa}
\end{subfigure}
\begin{subfigure}[b]{0.4\textwidth}
  \includegraphics[width=\linewidth]{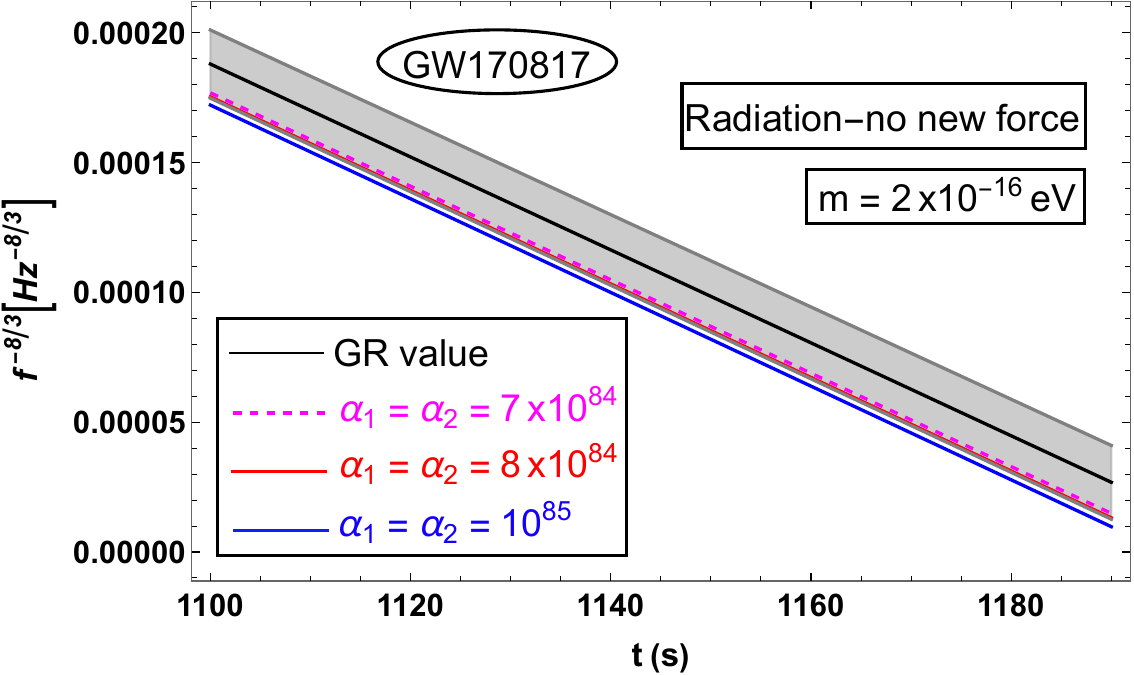}
  \caption{Constraints on couplings from radiation only for $m=2\times 10^{-16}~\mathrm{eV}$}
  \label{fig:gwtwob}
\end{subfigure}
\caption{Constraints on the couplings $\alpha_i$ from the GW170817 event in ghost-free quadratic gravity, including radiation effects only, shown for (a) $m=10^{-15}~~\mathrm{eV}$ and (b) $m=2\times 10^{-16}~~\mathrm{eV}$. See texts for details.}
\label{fig:gwtwo}
\end{figure}

In FIG. \ref{fig:gwtwo}, we present the evolution of the GW frequency as a function of the coalescence time for the GW170817 event in ghost-free quadratic gravity, considering only radiation effects without any additional force. This corresponds to the case where the mode masses satisfy $m \lesssim 2\Omega$, such that the exponential terms in Eq. \ref{rad10} do not contribute to the orbital frequency evolution. Results are shown for representative masses $m=10^{-15}~\mathrm{eV}$ (FIG. \ref{fig:gwtwoa}) and $m=2\times 10^{-16}~\mathrm{eV}$ (FIG. \ref{fig:gwtwob}).

From these cases, we derive bounds $\alpha_i \lesssim 3\times 10^{83}$ for $m = 10^{-15}~\mathrm{eV}$ and $\alpha_i \lesssim 8\times 10^{84}$ for $m = 2\times 10^{-16}~\mathrm{eV}$, indicated by the red curves. The blue curves correspond to $\alpha_i=4\times 10^{83}\,(10^{85})$ for $m=10^{-15}\,(2\times 10^{-16})~\mathrm{eV}$, which lie outside the chirp-mass measurement band, while the values $\alpha_i=2\times 10^{83}\,(7\times 10^{84})$ fall within it, which are shown by the magenta dashed lines. The black curve denotes the GR prediction.

The coalescence time is shifted by $\mathcal{O}(7~\mathrm{s})$ relative to GR for $\alpha_i=3\times 10^{83},\, m=10^{-15}~\mathrm{eV}$ and for $\alpha_i=8\times 10^{84},\, m=2\times 10^{-16}~\mathrm{eV}$. We find that the constraints become tighter as the mode mass increases, though the bounds obtained in the radiation-only case are weaker than those derived when new-force effects are included.

\subsubsection{Combined effects of both new force and radiation}
\begin{figure}[htbp]
\centering
\begin{subfigure}[b]{0.4\textwidth}
  \includegraphics[width=\linewidth]{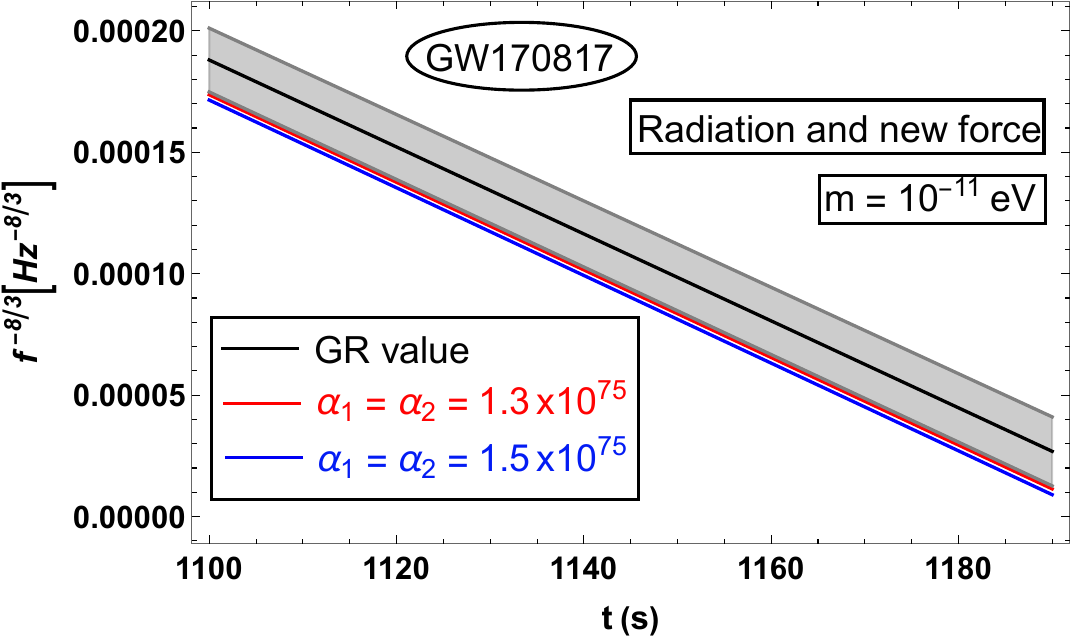}
  \caption{Constraints on couplings from new force+radiation effects for $m= 10^{-11}~\mathrm{eV}$}
  \label{fig:gwthreea}
\end{subfigure}
\begin{subfigure}[b]{0.4\textwidth}
  \includegraphics[width=\linewidth]{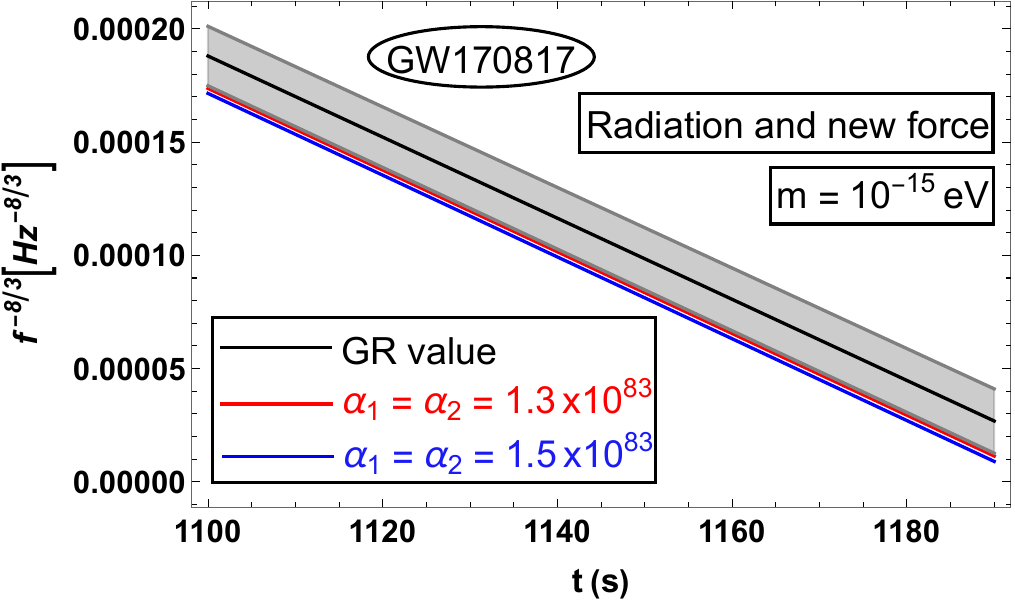}
  \caption{Constraints on couplings from new force+radiation effects for $m=10^{-15}~\mathrm{eV}$}
  \label{fig:gwthreeb}
\end{subfigure}
\caption{Constraints on the couplings $\alpha_i$ from the GW170817 event in ghost-free quadratic gravity including both radiation and new force effects, shown for $m= 10^{-11}~\mathrm{eV}$ and $m=10^{-15}~\mathrm{eV}$. See texts for details.}
\label{fig:gwthree}
\end{figure}

In FIG. \ref{fig:gwthree}, we show the evolution of the GW frequency with coalescence time in ghost-free quadratic gravity, including the combined effects of both the new force and massive-mode radiation. The results are obtained using Eq. \ref{rad10}, where contributions from both the exponential factors and the $(1-m^2/4\Omega^2)$ terms are taken into account. Illustrative cases are presented for mode masses $m=10^{-11}~\mathrm{eV}$ (FIG. \ref{fig:gwthreea}) and $m=10^{-15}~\mathrm{eV}$ (FIG. \ref{fig:gwthreeb}).

For $m= 10^{-11}~\mathrm{eV}$, we find $\alpha_i \lesssim 1.3\times 10^{75}$, which is slightly stronger than the limit from the new-force-only scenario (FIG. \ref{fig:gw1}). Likewise, for $m= 10^{-15}~\mathrm{eV}$, the constraint $\alpha_i \lesssim 1.3\times 10^{83}$ improves upon the radiation-only bound (FIG. \ref{fig:gwtwoa}). These limits are indicated by the red curves, while the blue curves corresponding to $\alpha_i=1.5\times 10^{75}\,(1.5\times 10^{83})$ at $m=10^{-11}\,(10^{-15})~\mathrm{eV}$ fall outside the observational uncertainty band. The inferred coalescence time is shifted by $\mathcal{O}(7.7~\mathrm{s})$ relative to GR for $\alpha_i=1.3\times 10^{75}$, $m=10^{-11}~\mathrm{eV}$, and for $\alpha_i=1.3\times 10^{83}$, $m=10^{-15}~\mathrm{eV}$. As expected, accounting for both new-force and radiation effects leads to stronger bounds than either contribution considered separately.

\section{Conclusions and discussions}\label{sec9}

GR is widely regarded as an effective low-energy theory of gravity. In the weak-field, infrared regime it agrees with observations to excellent precision. To explore gravity in the ultraviolet regime, one augments the Einstein-Hilbert action with higher-curvature invariants (e.g., $R^2$, $R_{\mu\nu}R^{\mu\nu}$, $\cdots$), but any extension must be vetted for theoretical consistency such as renormalizability and unitarity. The canonical quadratic (``fourth-order'') theory is power-counting renormalizable in $4D$ yet propagates a massive spin-$2$ ghost, threatening unitarity. Ghost-free constructions can be achieved, for example, by enlarging the geometric sector (torsion and non-metricity) or by suitable derivative structures that eliminate light ghosts. In this work we probe the infrared phenomenology of both the standard and ghost-free fourth-order gravity theories, deriving constraints on the couplings and masses from the orbital period decay of quasi-stable binaries and the GW170817 frequency-chirp (chirp-mass) measurements.

We derive the modified gravitational potential in standard and ghost-free fourth-order gravity and compute the GW energy loss due to emission of the massless spin-$2$ graviton and the massive spin-$2$ and spin-$0$ modes for both quasi-stable and coalescing binaries. The calculation uses a QFT framework in which the binary is treated as a classical effective one-body source in the center-of-mass frame, while the radiated modes are quantized fields. Consequently, the orbital period decay and orbital-frequency evolution receive corrections relative to GR from the altered force law and additional radiation channels. The relative impact of these effects depends on the mode masses, or equivalently on the characteristic length scales probed by the measurements.

In standard fourth-order gravity, the energy carried by massless gravitons in the infrared is exactly canceled by the combined contribution of the massive spin-$2$ ghost and the massive spin-0 scalar. In the limit of zero mode masses, the total radiated power therefore vanishes, and the leading nonzero contribution appears at order $\Omega^{4}$ for non-zero mode masses (in contrast to the GR quadrupole result, which scales as $\Omega^{6}$). This cancellation does not imply an absence of GW emission. It permits emission of positive and negative-energy wave pairs at no net energy cost to the source, which at the quantum level leads to a catastrophic vacuum instability.

We apply this framework to the quasi-stable binaries, Hulse-Taylor and PSR J1738+0333 and compute their orbital period decay. In practice, the new-force contribution is relevant when the mode mass satisfies $m \lesssim 1/a$ (set by the orbital separation $a$), while radiation into massive modes becomes kinematically allowed for $m \lesssim \Omega$ (the orbital frequency), for the fundamental mode.  Owing to the leading-order cancellation, no universal upper bound on the mode masses can be extracted for $m\lesssim \Omega$ or $m\lesssim 1/a$. Even if the ghost were effectively decoupled, the modified potential would shift the orbital period decay outside the observational window. Therefore, we only obtain lower bounds on the masses of the modes to be $m\gtrsim 6.2\times 10^{-16}~\mathrm{eV}$.

By contrast, in ghost-free quadratic gravity, the Newtonian potential and the quadrupole radiation formula are recovered as the couplings go to zero independent of the choice of the masses. From the combined effect of the modified force law and radiation we obtain $\alpha_{1}\simeq\alpha_{2}\lesssim 4.13\times 10^{83}$ at the characteristic scale $m_{0+}\sim m_{2+}\sim 3.1\times 10^{-16}\,\mathrm{eV}$; from radiation-only we find $\alpha_{1}\simeq\alpha_{2}\simeq 1.4\times 10^{89}$ at $m_{0+}\sim m_{2+}\sim 1.2\times 10^{-18}\,\mathrm{eV}$, from the orbital period loss of quasi-stable binaries. Below these scales the theory is observationally degenerate with GR, whereas above them the long-range (mediator) assumption no longer holds. The bounds on the coupling become stronger for larger mass of the modes.

We likewise extract constraints on the mass of the modes and couplings of standard and ghost-free fourth-order gravity from the GW170817 coalescence binary, which probes shorter length scales (the stars approach within $\sim 20~\mathrm{km}$) and higher orbital frequencies than quasi-stable binaries $(\sim 10~\mathrm{Hz})$. As in the quasi-stable case for standard fourth-order gravity, the chirp-mass measurement yields no universal bound on the mode masses at $\gtrsim 20~\mathrm{km}$ because the leading contributions to $\dot{\Omega}$ cancel among modes. The first non-vanishing term scales as $\Omega^{5/3}$, in contrast to GR where $\dot{\Omega}\propto \Omega^{11/3}$. Therefore, we obtain the lower bound on the mass parameters for the ghostful theory as $m\gtrsim 10^{-11}~\mathrm{eV}$ from GW170817 event. In ghost-free quadratic gravity, by contrast, the GR behavior is recovered in the limit of vanishing couplings. From the combined new-force and radiation effects we obtain our strongest coupling limit, $\alpha_1\simeq\alpha_2 \lesssim 1.3\times 10^{75}$, at the characteristic scale $m_{0+}\sim m_{2+}\sim 10^{-11}~\mathrm{eV}$.

The constraints on the couplings are inherently scale dependent: as the observational length scale decreases (or equivalently, as the characteristic mode mass increases), the bounds on the couplings become increasingly stringent. Our analysis shows that GW observations disfavor the presence of ghost degrees of freedom. In particular, the standard fourth-order gravity model fails to reproduce the GR quadrupole formula in the weak-field limit, whereas the ghost-free quadratic gravity framework recovers it consistently. Beyond theoretical requirements such as renormalizability and unitarity, our results indicate that ghostful theories are also phenomenologically disfavored by current GW data. Even tighter constraints are expected as one probes shorter length scales approaching the Planck regime, although at such scales the UV completion is expected to deviate from the linearized framework adopted here, and nonlinear effects are likely to become significant. In the regime where the mass parameters greatly exceed the characteristic inverse length scale of the system, the corresponding Yukawa-type terms associated with massive modes in the potential are exponentially suppressed. Moreover, massive radiating modes are kinematically forbidden when their mass exceeds the orbital angular frequency for a fundamental mode of a quasi-stable orbit and two times the orbital frequency for inspiraling circular binary. Consequently, the contributions from all massive modes are strongly suppressed, leaving only the massless graviton as the dominant propagating degree of freedom.

Looking ahead, the analysis can be extended to binary BHs \cite{LIGOScientific:2016vpg}, mixed binaries \cite{Thompson:2020nei}, and extreme mass-ratio inspirals \cite{Yunes:2011aa}, with explicit waveform modeling of strain and phase modifications from extra modes. Post-merger/ringdown observations offer sensitivity to larger mode masses (higher GW frequencies) and thus a complementary window for constraints. The quantum nature of gravity can also be investigated through tabletop experiments, as discussed in \cite{Carney:2018ofe}. In particular, detecting single gravitons via spontaneous or stimulated emission processes would serve as a definitive signature of gravity's quantization. Several experimental proposals aimed at probing these effects have been discussed in \cite{Tobar:2023ksi,Carney:2023nzz}. It is also natural to compare with ghost-free nonlocal gravity, where vacuum pathologies are avoided and GW energy-loss tests and modified force law have already yielded bounds at different length scales \cite{Calmet:2016sba,Calmet:2017rxl,Calmet:2018hfb,Calmet:2018rkj,Capozziello:2024bjb,Ewasiuk:2025uwm}. Finally, ultralight massive modes in these theories may leave signatures across astrophysics, providing additional avenues to constrain these theories with forthcoming multi-band GW and multi-messenger data. We will analyze these signatures in separate publications.

\section*{Acknowledgements}
This article is based on the work from COST Actions COSMIC WISPers CA21106 and BridgeQG CA23130, supported by COST (European Cooperation in Science and Technology). 
The work of S.M. was supported in part by the Japan Society for the Promotion of Science (JSPS) Grants-in-Aid for Scientific Research No.~24K07017 and the World Premier International Research Center Initiative (WPI), MEXT, Japan. T.K.P would like to thank ICTP for the kind hospitality during the finalization of the paper. 

\bibliographystyle{utphys}
\bibliography{reference}
\end{document}